\documentclass{eas}
\usepackage{graphicx}
\usepackage{amssymb}
\usepackage{rotating}

\usepackage[sectionbib]{natbib}
\bibpunct{(}{)}{;}{a}{}{,}
\begin{document}

\title{Detection and Characterization of Extrasolar Planets through Doppler
Spectroscopy}\thanks{We warmly thank the organizers for their invitation. We
thank Olivier Preis for comments that helped improve the quality of the 
manuscript. A.~E. acknowledges support from the French National Research Agency  
through project grant ANR-NT-05-4\_44463.}
\runningtitle{Detection and Characterization of Extrasolar Planets}
\author{A. Eggenberger}
\address{Laboratoire d'Astrophysique de Grenoble, 
UMR 5571 CNRS/Universit\'e Joseph Fourier, BP~53, F-38041 Grenoble Cedex 9, 
France; \email{Anne.Eggenberger@obs.ujf-grenoble.fr}}
\author{S. Udry}
\address{Observatoire de Gen\`eve, 
Universit\'e de Gen\`eve, 51 ch. des Maillettes, CH-1290 Sauverny, Switzerland; 
\email{Stephane.Udry@unige.ch}}
\begin{abstract}
Over 300 extrasolar planets have been found since 1992, showing that 
planetary systems are common and exhibit an outstanding variety of  
characteristics. 
As the number of detections grows and as models of planet formation 
progress to account for the existence of these new worlds, statistical 
studies and confrontations of observation with theory allow to 
progressively unravel the key processes underlying planet formation. 
In this chapter we review the dominant contribution of Doppler
spectroscopy to the present discoveries and to our general understanding of
planetary systems. We also emphasize the synergy of Doppler 
spectroscopy and transit photometry in characterizing the physical properties of
transiting extrasolar planets. As we will see, Doppler spectroscopy has not 
reached its limits yet and it will undoubtly play a leading role in the 
detection and characterization of the first Earth-mass planets.
\end{abstract}
\maketitle

\section{Introduction}
\label{intro}

The question of the existence of other worlds has been present in human
history for millennia but it is only recently that scientific evidence
has confirmed what many had anticipated: planets do exist and are common 
outside the Solar System. The first robust detection of another planetary 
system came in 1992 with the discovery of two terrestrial-mass planets orbiting 
the pulsar PSR\,1257+12 \citep{Wolszczan92}. Interestingly, this 
discovery did not receive all the attention that could have been 
expected, probably because these two planets orbit a 
``dead star'' very different from the Sun and much less likely to host life in
its vicinity. From this anthropocentric perspective, the major 
milestone in the search for extrasolar
planets was the discovery in 1995 of 51\,Peg\,b, the first extrasolar planet 
found to orbit a Sun-like star \citep{Mayor95}. Although of Jovian nature, 
51\,Peg\,b orbits at 0.052 AU from its parent star, a striking characteristic
when compared to the giant planets in the Solar System, which all 
orbit beyond 5 AU. This proximity has been a major surprise 
and a serious challenge to planet formation theories.

Thirteen years after the discovery of 51\,Peg\,b, over 300 extrasolar
planets have been detected, including many fascinating 
systems\footnote{See the Extrasolar Planets Encyclopedia, 
http://exoplanet.eu/, for an up-to-date census of the discoveries.}. 
Five different techniques have contributed to these discoveries: 
pulsar timing \citep[4 planets detected; e.g.][]{Wolszczan97}, 
Doppler spectroscopy \citep[292 planets; e.g.][]{Udry07}, 
photometric transits 
\citep[52 planets; e.g.][]{Charbonneau07}, 
microlensing \citep[8 planets; e.g.][]{Gaudi07}, and direct imaging 
\citep[4 objects with a mass possibly below 20 M$_{\rm Jup}$; e.g.][]{Beuzit07}. 
These observational techniques have considerably improved in recent years and 
Doppler spectroscopy, which has contributed the bulk of the discoveries so far, 
now allows the detection of planets with half the mass of Uranus.

Likewise, since the discovery of 51\,Peg\,b, planet formation theory has made 
rapid progress. To overcome the challenge posed by the unexpected properties 
of the newly found planets, two different formation models 
have been proposed: core accretion and disk instability. 
According to the core accretion model, dust grains coagulate to form
planetesimals, which then accumulate to build up planetary cores. Planetary
cores reaching the critical mass of 5-15~M$_{\oplus}$ before the 
dissipation of the gaseous protoplanetary disk subsequently accrete significant 
amounts of nebular gas and become giant planets \citep[e.g.][]{Lissauer07}. 
The remaining solid cores merge through giant impacts to form terrestrial 
planets \citep[e.g.][]{Nagasawa07}. In the alternative disk instability 
model, giant planets form by direct fragmentation of the protoplanetary disk 
\citep[e.g.][]{Durisen07}. 
Quantitative predictions based on the disk instability
scenario are still sparse because the simulations are computationally 
challenging and involve complex physics. In contrast, core accretion is
now mature enough to allow for detailed calculations, and explicit comparisons
with the observed population of extrasolar planets are possible.

In this chapter we review the dominant contribution of Doppler spectroscopy
to planet discoveries and to our general understanding of planetary systems. In
Sect.~\ref{doppler_spectroscopy} we describe the Doppler technique itself. 
Then, we present the observational results and we discuss their interpretation 
within the current theoretical framework. In Sect.~\ref{stat_giants} we
consider the results on giant planets for which an extended statistics is 
available. In Sect.~\ref{light_planets} we present recent results on 
low-mass planets of Neptune- and Earth-type, and we discuss the present 
limitations on Doppler precision. 
We end this review by describing in Sect.~\ref{transits} the role 
played by Doppler spectroscopy in the characterization of transiting planets. 
All these achievements and results are summarized in Sect.~\ref{conclusion}, 
where we also outline future perspectives.


\section{Doppler spectroscopy}
\label{doppler_spectroscopy}

\subsection{Principle}
\label{principle_doppler}

\begin{figure}
\centering
\resizebox{0.9\textwidth}{!}{
\includegraphics{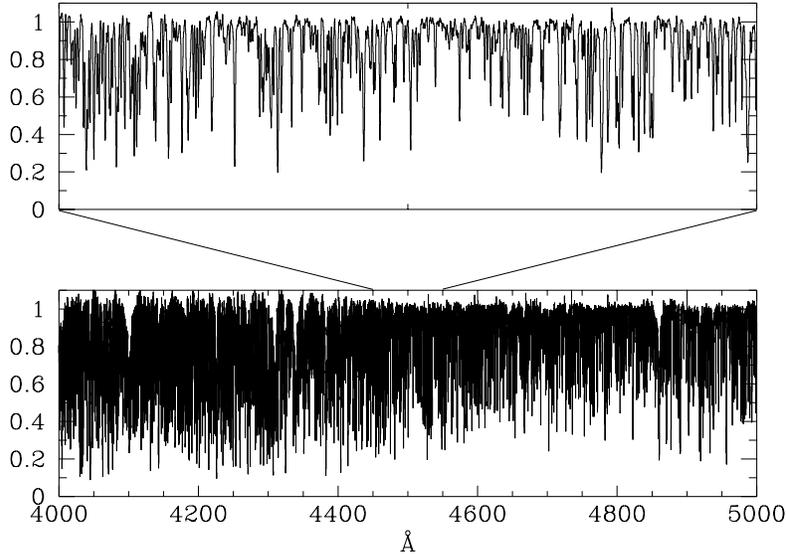}
}
\caption{Blue part of an ELODIE spectrum of the planet-host star 51 Peg
illustrating the multitude of absorption lines present in the visible spectra 
of solar-type stars.}
\label{51peg_elodie_spectrum}
\end{figure}

Doppler spectroscopy is an indirect
detection method which uses the starlight to measure the gravitational 
influence of a planet on its host star.
Specifically, Doppler spectroscopy is based on the following key observations:
\begin{enumerate}
\item In a planetary system, the star and the planet orbit their common 
barycenter according to Newton's law of gravitation and to the laws of motion. 
The two barycentric and the relative orbits have the same periods and 
eccentricities, but semimajor axes in the proportions 
$a_{\star}$\,:\,$a_p$\,:\,$a$\,$=$\,$m_p$\,:\,$m_{\star}$\,:\,$(m_{\star}+m_p)$, where 
$m_{\star}$ is the mass of the parent star and $m_p$ the mass of the planet. 
The three orbits are coplanar and the orientations of the two barycentric 
orbits differ by $180^{\circ}$ within that plane.
\item According to the Doppler-Fizeau effect (hereafter simply the Doppler 
effect), the light emitted by a source approaching (receeding from) 
the observer is shifted towards shorter (longer) wavelengths. In its simplest
relativist form, the Doppler formula writes
\begin{eqnarray}
z = \frac{\lambda - \lambda_0}{\lambda_0} = \frac{1 + V_r/c}{\sqrt{1-V^2/c^2}} - 1
\label{eq0}
\end{eqnarray}
where $z$ is the so-called redshift, $\lambda$ and $\lambda_0$ are the observed
and rest wavelengths, respectively, $V_r$ is the radial velocity, and $V$ is the
total velocity relative to the observer. In the $\ln{\lambda}$ space, the 
wavelength shift is independent of the rest wavelength and provides a direct
measurement of the relative velocity between the source and the observer. 
\item The visible portion of the spectra of F, G, K and M dwarfs contains a
multitude of metal absorption lines (Fig.~\ref{51peg_elodie_spectrum}). 
These lines constitute a convenient 
benchmark to measure wavelength shifts through the Doppler effect. 
\end{enumerate}

Putting these three observations together, Doppler spectroscopy consists in 
monitoring potential Doppler shifts in the spectra of stars with numerous 
absorption lines. In the absence of other phenomena susceptible of shifting the 
stellar lines or of modifying their profile (Sect.~\ref{astro_limitations}), 
measured Doppler shifts are converted into radial (line-of-sight) velocity 
variations and interpreted as the reflex motion of the star due to an orbiting 
companion (planet, brown dwarf, or star). Since the two barycentric and the 
relative orbits are closely related, measuring the reflex motion of the parent 
star along the line of sight gives access to some of the orbital parameters and 
characteristics of the companion.

\subsection{Planetary orbits and planet characteristics from radial velocities}
\label{kep_analysis}

\begin{figure}
\centering
\resizebox{\textwidth}{!}{
\includegraphics[115,109][530,712]{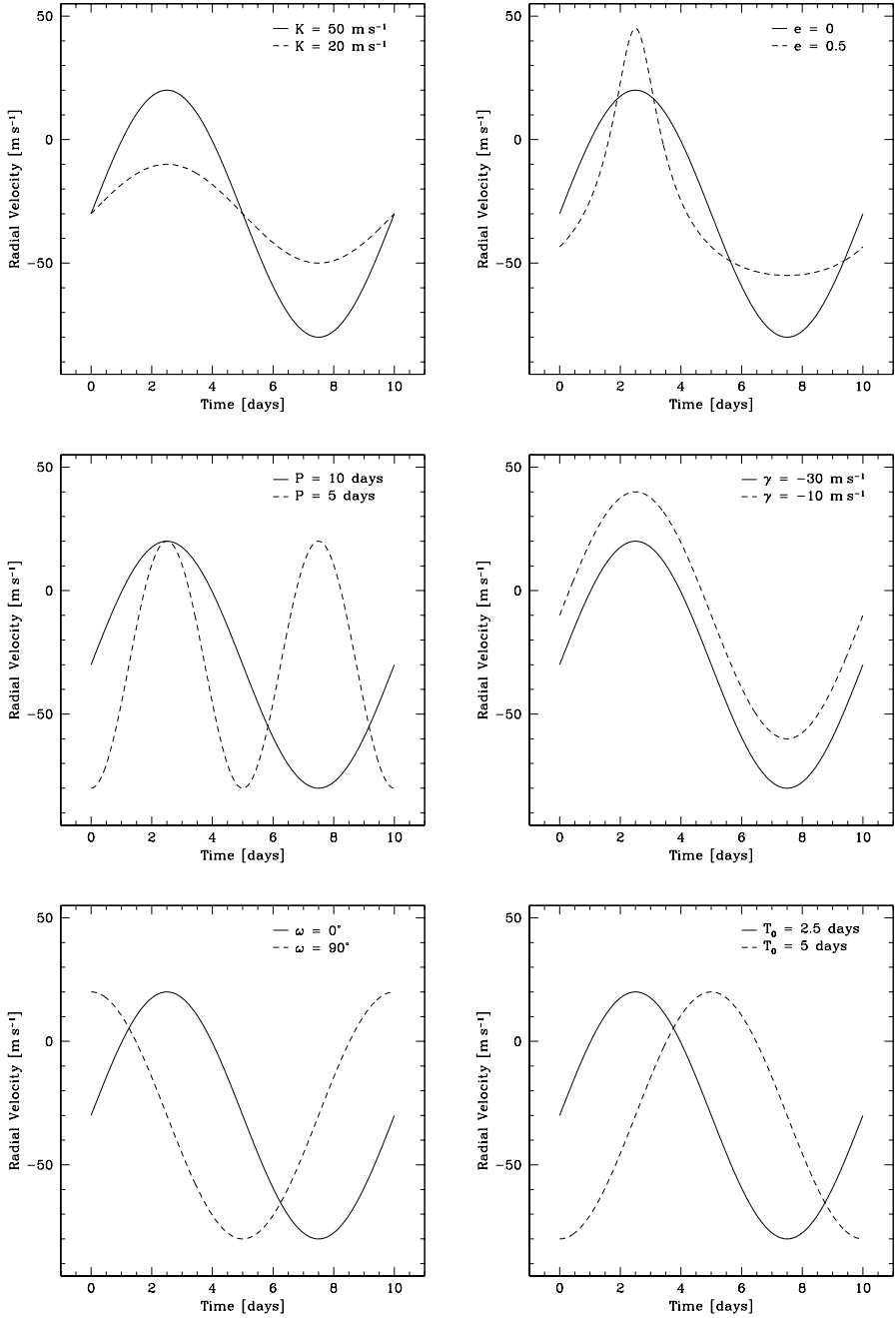}
}
\caption{Illustration of the influence of each of the six parameters determining
the Keplerian model of Eq.~\ref{eq1}. The baseline model has $P=10$ days, $e=0$,
$K=50$ m\,s$^{-1}$, $\omega = 0^{\circ}$, $T_0 = 2.5$ days, and 
$\gamma = -30$ m\,s$^{-1}$. These orbital elements are varied in turn, one on 
each panel.}
\label{spectro_elements}
\end{figure}

A planet in a Keplerian orbit induces on its parent star a perturbation of the 
form
\begin{eqnarray}
V_r(t) = K\,[\cos(\nu(t) + \omega) + e\cos(\omega)] + \gamma 
\label{eq1}
\end{eqnarray}
where $K$ is the velocity semiamplitude 
\begin{eqnarray}
 K = \frac{2 \pi a_{\star} \sin{i}}{P{(1-e^2)}^{1/2}}
\label{eq2}
\end{eqnarray}
$\omega$ is the longitude of periastron, and $\gamma$ is the systemic
velocity (velocity of the barycenter). Since the true
anomaly, $\nu(t)$, depends on the orbital period ($P$), eccentricity ($e$) and 
time of passage at periastron ($T_0$), fitting a radial-velocity time series with 
the Keplerian model described above yields six parameters: 
$K$, $e$, $w$, $T_0$, $P$, and $\gamma$ (Fig.~\ref{spectro_elements}).

For a planetary system, the velocity semiamplitude is related to the masses of 
the two components through the so-called mass function 
\begin{eqnarray}
\frac{{(m_p\sin{i})}^3}{(m_{\star}+m_p)^2} = \frac{P}{2 \pi G}\,K^3\,{(1-e^2)}^{3/2}
\label{eq3}
\end{eqnarray}
which reduces to the expression of the planet minimum mass 
\begin{eqnarray}
m_p\sin{i} \simeq \left(\frac{P}{2 \pi G}\right)^{1/3}\,K\,m_{\star}^{2/3}\,{(1-e^2)}^{1/2}
\label{eq4}
\end{eqnarray}
under the assumption that $m_p \ll m_{\star}$ 
($G$ is the universal gravitational constant). Introducing the same
approximation into Kepler's third law yields an expression for the semimajor 
axis of the relative orbit
\begin{eqnarray}
a \simeq a_p \simeq \left(\frac{G}{4 \pi^2}\right)^{1/3} m_{\star}^{1/3} P^{2/3}
\label{eq4b}
\end{eqnarray}
Therefore, fitting radial-velocity data with a Keplerian model to account for
the presence of a planetary companion gives four of the six orbital elements of 
the relative orbit (the longitude of the ascending node, $\Omega$, and the 
orbital inclination, $i$, remain unknown). When the mass of the central star can be estimated by other 
means, the Keplerian fit additionally yields a lower limit on the planetary 
mass and the semimajor axis of the relative orbit. Although the true mass of 
the planet can be significantly different from the minimum mass, the two values 
agree within a factor of 2 ($m_p \leq 2\,m_p\sin{i}$) in 87\% of cases.

For a planet in a circular orbit around a solar-mass star, Eq.~\ref{eq4} 
simplifies to 
\begin{eqnarray}
K\,[\mathrm{m}\,\mathrm{s}^{-1}] \simeq 28.57\ m_p\sin{i}\,[\mathrm{M}_{\mathrm{Jup}}]\ P^{-1/3}\,[\mathrm{yr}]
\label{eq5}
\end{eqnarray}
Since the probability of detecting a planetary signal depends essentially on 
the value of the velocity semiamplitude, this expression indicates that 
Doppler measurements favor the detection of planetary systems with massive and  
short-period planets. Applied to the Solar System, Eq.~\ref{eq5} shows that 
Jupiter induces on the Sun a radial-velocity perturbation with a semiamplitude of 
12.5 m\,s$^{-1}$, while the Earth induces a perturbation with a semiamplitude 
of 9 cm\,s$^{-1}$. As we will discuss now, obtaining radial 
velocities at this level of precision is nontrivial.

\subsection{Instrumentation and techniques for high-precision Doppler measurements}

A measurement error of 1 m\,s$^{-1}$ corresponds to 
$\Delta \lambda/\lambda_0$\,$=$\,$3$\,$\times$\,$10^{-9}$. 
On the detector of a typical 
spectrograph ($R$\,$\sim$\,100,000), this wavelength shift 
translates into a linear displacement of about $3$\,$\times$\,$10^{-4}$ 
resolution element ($\sim$$10^{-4}$ line width). 
Although attaining this precision is feasible by averaging over many  
spectral lines, experience shows that unless one
takes special measures, errors in Doppler measurements are much greater than 
the above values and are usually systematic \citep{Griffin73,Brown90}.

\subsubsection{Requirements for high-precision Doppler measurements}
\label{systematics}

Doppler measurements made with spectrometers or classical spectrographs 
rarely achieve standard errors better
than $\sim$200 m\,s$^{-1}$ \citep[e.g.][]{Marcy89,Duquennoy91a}, 
which falls short of what is needed to detect extrasolar planets. 
To achieve the high Doppler precision necessary for the 
detection of extrasolar planets, the following 
sources of systematic errors must be addressed:

\begin{enumerate}
\item Motion of the photocenter at the spectrograph slit. 
In Doppler units, the slit width of a spectrograph is typically 
a few km\,s$^{-1}$, meaning that errors larger than 1 m\,s$^{-1}$ will occur if 
the photocenter moves away from the slit center by a few $10^{-3}$ slit widths. 
In practice, photocenter motions due to guiding errors, focus, seeing fluctuations, 
and atmospheric refraction usually amount to $\sim$$10^{-1}$ slit widths. 
An efficient solution to this problem is to use an optical fiber (often with the 
addition of a scrambling device) to convey and scramble the 
starlight from the telescope to the spectrograph, thereby producing a nearly 
uniformly illuminated disk at the spectrograph entrance. 

\item Variations of air refractive index and thermomechanical flexures. 
Variations of temperature and barometric pressure modify the refractive index 
of air near the grating, causing spurious wavelength shifts similar to Doppler
shifts. Mechanical instabilities and thermal relaxation also produce 
non-negligible motions of the spectrum relative to the detector, causing 
additional spurious shifts and changes in the instrumental profile. 
For the CORALIE spectrograph, a temperature change of 1~K 
produces a net velocity drift of $\sim$90~m\,s$^{-1}$, 
while a pressure change of 1~mbar produces a net velocity drift of 
$\sim$300 m\,s$^{-1}$. A partial solution to this problem is to stabilize and 
control the entire spectrograph in temperature and pressure. 
Yet, small wavelength shifts cannot be fully avoided and the 
general way to deal with this problem is to use a simultaneous 
wavelength calibration to monitor -- and then correct for -- these 
``instrumental drifts''.

\item Timing of exposure and barycentric correction. 
Classical radial velocities are corrected to the Solar System
barycenter. The main contributions to the Earth's barycentric motion are
the diurnal rotation of the Earth (1-2 m\,s$^{-1}$ per minute at most) and 
the Earth's orbital revolution ($\pm$30 km\,s$^{-1}$ per year). Yet, at 
the m\,s$^{-1}$ level, the motion of the Sun around the Solar System barycenter and the motion 
of the Earth around the Earth-Moon barycenter are also significant. To obtain 
precise barycentric radial velocities, one thus 
needs precise Solar System ephemeris and one needs to know the photon-weighted 
midpoint of each observation to better than 30~s.
\end{enumerate}

The above requirements show that the secret to making high-precision Doppler
measurements is to record the wavelength reference spectrum simultaneously
with the stellar spectrum and to efficiently scramble the starlight before
sending it to the spectrograph. In terms of measurements, the observed 
radial velocity ($V_{\rm obs}$) must be corrected for the instrumental drift 
($V_{\rm inst}$, mainly due to points 1 and 2 above) and for the Earth's 
barycentric motion ($V_{\rm Earth}$, point 3 above) 
to become the stellar barycentric radial velocity 
$V_{\rm star} = V_{\rm obs} - V_{\rm inst} - V_{\rm Earth}$ 
used in the Keplerian analysis described in Sect.~\ref{kep_analysis}. Two 
techniques have successfully been developed to accomplish this. If 
both these techniques use a cross-dispersed echelle spectrograph and 
track instrumental drifts by means of a
simultaneous wavelength calibration, they differ fundamentally in their 
approach and in their design.

\subsubsection{The simultaneous reference technique}

\begin{figure}
\centering
\resizebox{\textwidth}{!}{
\includegraphics{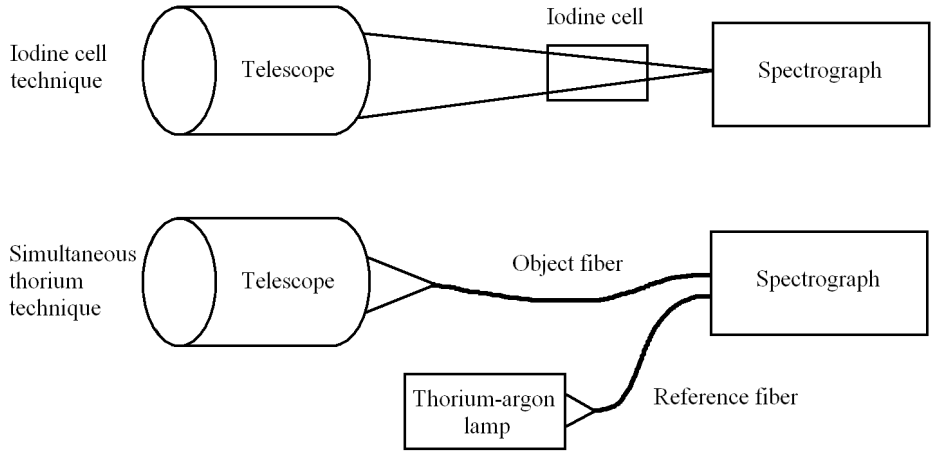}
}
\caption{Principle of the two techniques used to obtain high-precision radial 
velocities.}
\label{two_vrtechniques}
\end{figure}

As illustrated on Fig.~\ref{two_vrtechniques} (lower panel), the simultaneous 
reference technique involves 
the use of two optical fibers to feed the spectrograph: the so-called object 
fiber which records the 
starlight, and the so-called reference fiber which records the light 
from a wavelength reference source. In practice, the two fibers are brought into the 
entrance plane of the spectrograph in close proximity to one another, separated 
in the direction perpendicular to the main  dispersion. The object and 
the reference spectra are then recorded simultaneously, the two
sets of echelle orders being distinct and alternate on the detector. 
The simultaneous reference technique was pioneered on the ELODIE 
spectrograph \citep{Baranne96}, and up to now all the spectrographs based on 
this technique have used as reference source a thorium-argon (ThAr) lamp. The 
method is thus commonly known as the simultaneous thorium technique.

A distinguishing feature of the simultaneous reference technique is that it 
can achieve high radial-velocity precision only if implemented on fiber-fed 
spectrographs with high mechanical and thermal stability. 
The use of optical fibers equipped with a scrambling device is indeed 
essential to reduce instabilities in illumination and variations of instrumental
profile, both of which cannot be corrected by the simultaneous calibration. 
As to the thermal stabilization, it is meant 
to keep the optical paths of the two beams very similar within the 
spectrograph. Under such circumstances, the residual instrumental drifts 
experienced by the two beams are highly correlated and the velocity drift 
measured by the reference channel can be used to correct the object channel. 

\begin{figure}
\centering
\resizebox{\textwidth}{!}{
\includegraphics[29,184][506,692]{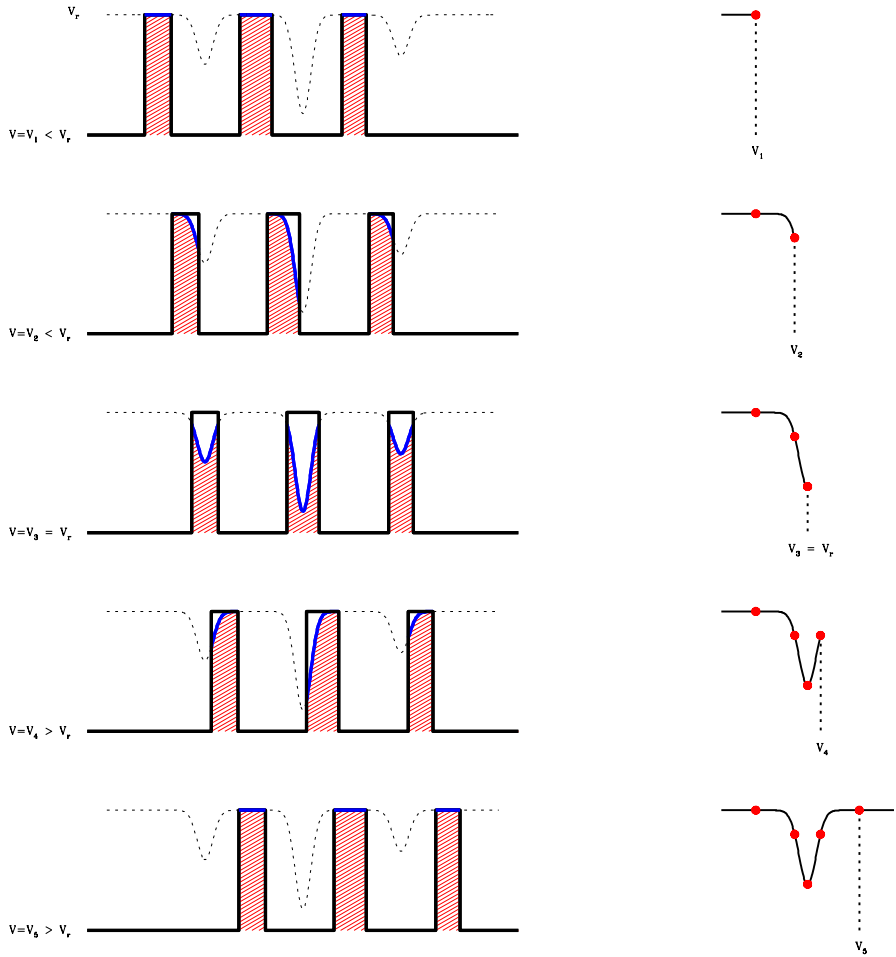}
}
\caption{Illustration of the construction of the cross-correlation function.
Diagrams on the left represent the stellar spectrum (dashed lines) and the 
binary mask (solid lines, transmission zones depicted as hatched 
areas). Diagrams on the right show the result of the cross-correlation process. 
Courtesy of Claudio Melo.}
\label{ccf_construction}
\end{figure}

If the simultaneous thorium technique puts considerable demands on the 
instrumentation, the Doppler analysis is relatively straightforward. 
Radial velocities are traditionally obtained by numerically cross-correlating 
the observed spectra with box-shaped, binary (0 and 1 values) templates called 
masks \citep{Baranne96,Pepe02b}. Schematically, the cross-correlation 
function (CCF) is constructed by shifting the velocity of the mask by increasing
amounts over a window roughly centered on the radial velocity of the star 
(Fig.~\ref{ccf_construction}). The
better the alignment between the stellar lines and their box-shaped 
counterparts in the mask, the lower the cross-correlation value. The CCF is
thus minimal when the velocity of the mask perfectly matches the radial 
velocity of the star (middle panel in Fig.~\ref{ccf_construction}). For slowly 
rotating dwarfs ($v\sin{i} \lesssim 10$ km\,s$^{-1}$), the central part of the 
CCF is well approximated by a Gaussian function and the radial velocity of
the star is measured by the center of the best-fit Gaussian.
Note that radial velocities obtained through cross-correlation are free from 
most of the systematic errors listed in Sect.~\ref{systematics} only relative 
to the stellar mask, which defines the velocity zero point.

The masks used for cross-correlation are of two types: stellar and thorium. 
Stellar 
spectra are cross-correlated with stellar masks whose nonzero zones corresponds 
to the theoretical positions and widths of stellar absorption lines at zero 
velocity. Experience shows that there is no need to use a different mask for 
each star. Common practice is to use a few masks 
corresponding to the main spectral subtypes (e.g. G2, K5, and M2). 
Stellar masks are built either from high-resolution, high signal-to-noise 
observed spectra or from synthetic spectra. Reference thorium spectra are 
cross-correlated with a 
thorium mask built from the atlas of \citet{Palmer83} and from the updated  
line list of \citet{Lovis07b}.

Stabilized spectrographs searching for planets using the simultaneous
thorium technique include 
ELODIE (1.93-m Telescope, Haute-Provence, France; dismounted in 2006), 
CORALIE (Leonard Euler Telescope, La Silla, Chile), HARPS (3.6-m ESO Telescope,
La Silla, Chile), and SOPHIE (1.93-m Telescope, Haute-Provence, France). The
radial-velocity precision has considerably improved from ELODIE to HARPS and is 
now better than 1 m\,s$^{-1}$ (Sect.~\ref{harps_limitations}).

\subsubsection{The gas cell technique}

The basic feature of the gas cell technique is to pass the starlight through a 
cell containing an absorbing medium (the reference source) before entry into 
the spectrograph (Fig.~\ref{two_vrtechniques}, upper panel). 
In this way, absorption lines from the reference source 
superimpose on the stellar spectrum, providing a fiducial wavelength 
scale that experiences the same instrumental shifts and distortions as the 
stellar spectrum. Since the gas cell is at rest relative to the observatory, 
spurious instrumental shifts are measured by the wavelength 
shift of the reference lines. The reference spectrum also provides a 
specification of the instrumental profile (IP) at each position on the 
detector, allowing for the measurement and correction of IP variations. 
The first application of the gas cell technique to planet 
searches was made by \citet{Campbell88}, who used a hydrogen fluoride (HF) 
cell \citep{Campbell79}. At present, all planet search programs based
on this technique use an iodine cell and the technique is commonly referred to 
as the iodine cell technique.

Instrumentally, the iodine cell technique is easily 
implemented on any existing slit spectrograph. The main complication of the 
method resides in the Doppler analysis. In practice, spectra taken through the 
iodine cell are broken up into several hundred ``chunks'' of length 
$\sim$2\ \AA. On each chunk the composite spectrum is modeled as 
\begin{eqnarray}
I_{obs}(\lambda) = k\left[T_{I_2}(\lambda)I_s(\lambda+\Delta
\lambda)\right]\ast IP
\label{eq8}
\end{eqnarray}
where $I_s$ is the intrinsic stellar spectrum, $T_{I_2}$ is the transmission
function of the iodine cell, IP is the in situ instantaneous instrumental 
profile, and the constant $k$ is a normalization factor. The wavelength shift, 
$\Delta\lambda$, is the topocentric Doppler shift of the star. The barycentric
radial velocity is obtained by correcting this topocentric Doppler shift for the
Earth's barycentric motion and by converting the result into a radial velocity
using an elaborated version of the Doppler formula. Again, the barycentric 
radial velocity obtained in this way is not an absolute velocity; it is precise 
only relative to a stellar template, which defines an arbitrary velocity 
zero point.

\begin{figure}
\centering
\resizebox{0.8\textwidth}{!}{
\includegraphics{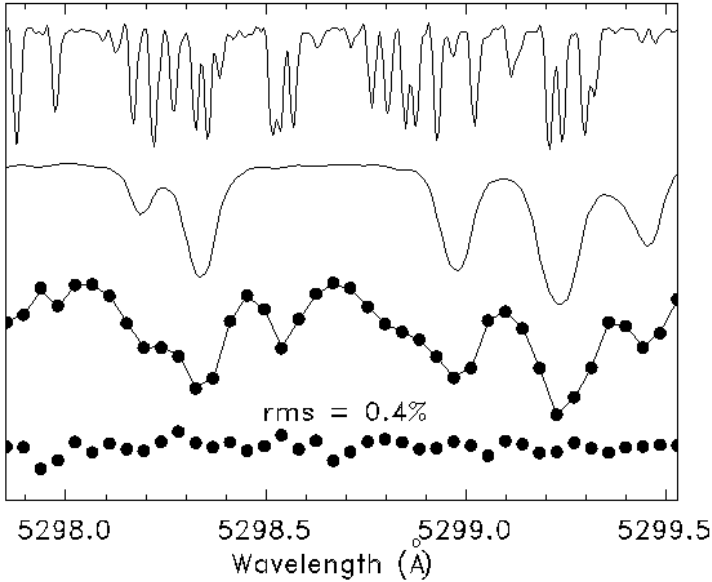}
}
\caption{Illustration of the modeling process for the iodine cell technique.
Top: transmission function of the iodine cell ($T_{I_2}$) 
as measured by the FTS. Second: a template stellar spectrum
(proxy for $I_s$). Third: the composite spectrum of the same
star, as observed through the cell (dots) and as modeled (solid line). 
Bottom: ten times the difference between the model and the
observations. Figure from \citet {Butler96}.}
\label{iodine_modeling}
\end{figure}

As shown by Eq.~\ref{eq8}, the modeling process requires two input functions, 
$I_s$ and $T_{I_2}$, plus the knowledge of the instrumental profile. 
The transmission function of the cell is measured directly
by obtaining a spectrum of the cell with a 
Fourier-Transform Spectrometer (FTS). The intrinsic stellar spectrum is more difficult
to obtain since it cannot be measured directly. $I_s$ is usually generated by 
taking a high signal-to-noise spectrum of the program star without the iodine 
cell in place (which gives $I_s$\,$\ast$\,$IP$) and by deconvolving this
spectrum from the instrumental profile. The instrumental profile cannot be
measured directly either. It is commonly generated by comparing observations of 
a hot, rapidly rotating star (which basically gives $T_{I_2}$\,$\ast$\,$IP$) 
with the reference FTS iodine spectrum. 

The whole modeling process is illustrated on Fig.~\ref{iodine_modeling} and
typically requires 13 parameters: 2 for the wavelength 
scale, 1 for the topocentric Doppler shift of the star, and 10 for the 
description of the instrumental profile. These parameters are determined by 
comparing a series of models to the observed spectrum through a least-squares 
fit. We refer the reader to \citet{Marcy92}, \citet{Valenti95}, and 
\citet{Butler96} for complementary information on the practical aspects of the 
iodine cell technique.

Echelle spectrographs equipped with an iodine cell and actively used for planet 
searches include HAMILTON (Shane/CAT, Lick, USA), HIRES (KECK, Mauna Kea, USA), 
UCLES (AAT, Siding Spring, Australia), HRS (HET, Davis
Mountains, USA), and UVES (VLT, Paranal, Chile). 
The iodine cell technique has long demonstrated a precision of 3 m\,s$^{-1}$
\citep{Butler96}, but the precision seems to have not much improved since then. 
Residuals around the best planetary orbits published so far indicate a 
precision of $\sim$2.5 m\,s$^{-1}$ \citep{Butler06,Fischer06,Wright08}.

\subsection{Limitations on Doppler spectroscopy}
\label{limitations}

Limitations on Doppler measurements can be classified into three broad 
categories: photon count, technical, and astrophysical. We describe here photon 
count and astrophysical limitations. Technical limitations related to the 
simultaneous thorium technique will be discussed in 
Sect.~\ref{harps_limitations}.

\subsubsection{Photon count limitations}

The ultimate limit to the attainable Doppler precision is that set by 
photon-counting statistics. As proposed by \citet{Connes85}, the uncertainty 
related to photon noise can be quantified by a quality factor, $Q$, which is a
sole function of the line profile in the case of pure photon noise. 
For spectral types between F2V and K7V, $Q$ increases towards the blue 
and $Q$ is the highest for a K5V star \citep{Bouchy01}. These two trends 
reflect the evolution in the strength and width of metal lines along 
the spectral sequence (late-type stars have more ``peaky'' lines) and the fact 
that these lines are more numerous 
between 4000 and 4500\ \AA\ than between 6000 and 6800\ \AA. $Q$ is 
also sensitive to rotational line broadening, 
decreasing as $1/v\,\sin{i}$ for $v\,\sin{i} > 6$ km\,s$^{-1}$. 
For these reasons, early-type stars ($\lesssim$F6) with few and broad 
absorption lines do not permit high-precision Doppler measurements, whereas 
slowly rotating K dwarfs are ideally suited for high-precision Doppler
programs.

\subsubsection{Astrophysical limitations}
\label{astro_limitations}

\paragraph{Stellar activity}

\begin{figure}
\centering
\resizebox{0.7\textwidth}{!}{
\includegraphics{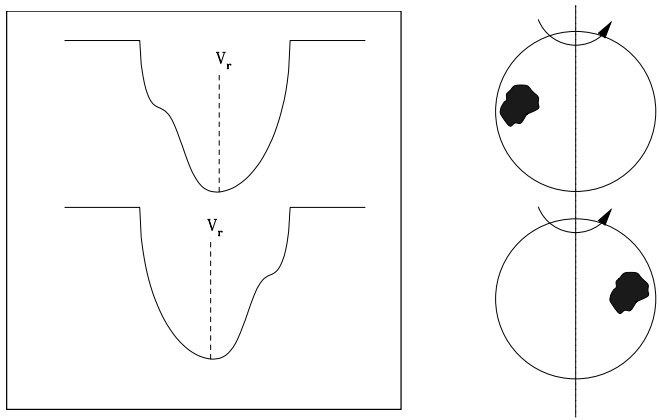}
}
\caption{Illustration of the effect of a starspot on the profile of the 
cross-correlation function and on the determination of its centroid (radial
velocity). The effect has been exaggerated for illustrating purposes.}
\label{spotted_star}
\end{figure}

Magnetic phenomena at the surface of solar-type stars induce radial-velocity
variations through the temporal and spatial evolution of spots, plages, and 
convective inhomogeneities \citep{Saar97,Saar98}. 
Figure~\ref{spotted_star} illustrates how the rotation of a starspot 
modifies the line profiles -- and equivalently the CCF profile -- affecting the
determination of the radial velocity. When the starspot pattern is long-lived, 
variations in line asymmetry are modulated by the rotational period of 
the star and can mimic a planetary signal \citep[e.g.][]{Queloz01,Bonfils07}. 
When 
the star is observed longer than the typical lifetime of starspots, the signal 
becomes incoherent and is detected as radial-velocity ``noise''. 
Although less confusing, this is still annoying since it may inhibit the 
detection of planetary signals of lower amplitude. 

In practice, radial-velocity perturbations related to stellar activity are
referred to as ``stellar jitter''. Stellar jitter depends on 
effective temperature, stellar activity, and projected rotational 
velocity \citep{Saar98,Santos00a,Wright05}, but these dependencies
cannot be modeled in detail yet. Typical values of stellar jitter are 
$\lesssim$\,5 m\,s$^{-1}$ for slowly rotating, chromospherically quiet G-K dwarfs 
\citep{Santos00a,Wright05} and $\lesssim$\,50 m\,s$^{-1}$ for F5-M2 young, 
active dwarfs \citep{Paulson04}. The above values explain why high-precision Doppler 
planet searches select their targets among old, slowly rotating inactive stars. 

To quantify the activity level of their (potential) targets, Doppler planet
searches traditionally use the parameter 
$R^{\prime}_{\rm HK}$, which represents the fraction of a star's 
bolometric flux emitted by the chromosphere in the Ca~\textsc{ii} H and K lines 
\citep{Noyes84}. Since this chromospheric emission is closely related to the
surface magnetic flux, a high $R^{\prime}_{\rm HK}$ value is an indication that 
a star may exhibit significant activity-related velocity variations. But  
a high $R^{\prime}_{\rm HK}$ index does not prove that a 
given Doppler signal is of stellar origin. To prove this, one can search for 
a periodic modulation in the Ca~\textsc{ii} flux, in the bisector of the
cross-correlation function, or in photometric observations 
\citep[e.g.][]{Queloz01,Desort07,Bonfils07}. These diagnostics are pretty
effective at identifying activity-related velocity variations in G, K, and M 
dwarfs.

\paragraph{Solar-like oscillations}
\label{oscillations}

\begin{figure}
\centering
\resizebox{0.8\textwidth}{!}{
\includegraphics{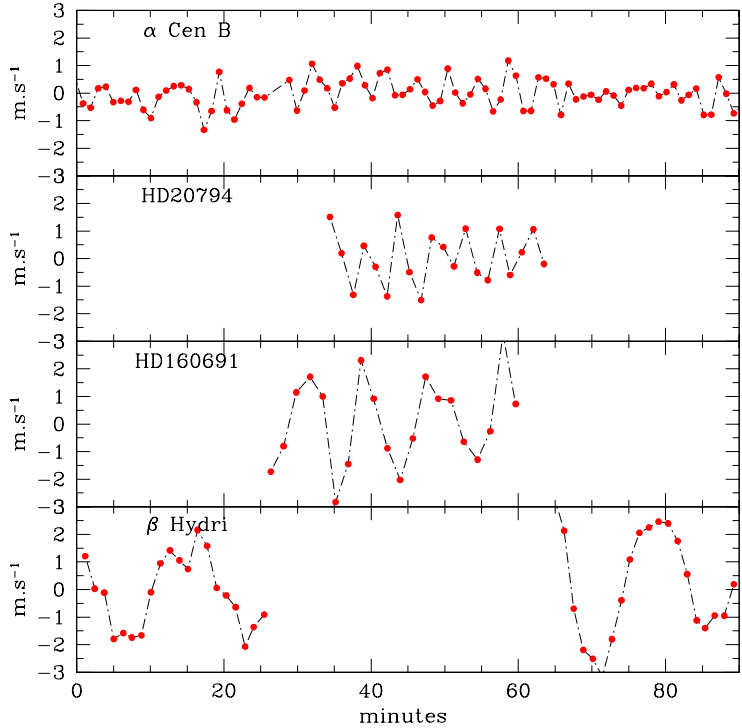}
}
\caption{Sequences of HARPS radial-velocity measurements on four solar-type
stars. Solar-like oscillations are clearly visible on each of these sequences.}
\label{solar_oscillations}
\end{figure}

The Sun oscillates in many low-amplitude modes simultaneously, each mode being
characterized by a maximum amplitude $\lesssim$\,20 cm\,s$^{-1}$ and a period 
between 3 and 15 minutes. This so-called 
five-minute oscillation is excited stochastically by turbulent convective 
motions near the solar surface. 
Main-sequence stars with significant outer convection zones should all exhibit 
similar pulsations, called solar-like oscillations in this more 
general context. Recent observations have shown that this is indeed the case 
(Fig.~\ref{solar_oscillations}). It should be emphasized that while individual 
modes have typical amplitudes in the range 10-50 cm\,s$^{-1}$ for main-sequence
stars, their interference can produce radial-velocity variations up to about 10 
times these values on time scales of minutes. Although these  
oscillations constitute a very interesting tool to probe the structure of 
stellar interiors (this is asteroseismology; see the contribution by Vauclair), 
they are essentially ``an annoying source of 
noise'' when viewed from the side of very high precision planet searches.
As explained in Sect.~\ref{harps_limitations}, an adequate observing strategy allows fortunately 
to largely average out this noise, reducing its impact on the search for 
low-mass planets.

\paragraph{Contamination effects}
\label{contamination}

Slit spectrographs used for planet searches have typical on-sky slit widths of 
$0.5$-$1^{\prime\prime}$, while fiber spectrographs have typical on-sky fiber 
diameters of $1$-$3^{\prime\prime}$. The light from an object close to the 
scientific target may thus also fall within the slit/fiber, contaminating the
target's spectrum and altering the determination of the target's velocity. 
For this reason, known binary stars closer than $2$-$6^{\prime\prime}$ are 
generally rejected from Doppler planet search programs 
(see \citeauthor{Eggenberger07}~\citeyear{Eggenberger07} for details). 
Yet, some close star systems are present in the samples of 
large-scale Doppler surveys and it may happen that a star turns out to be a 
binary, or that a presumed binary system turns out to be triple. The latter 
case is of particular concern because 
unrecognized triple systems may under some circumstances mimic the
radial-velocity signature of a planetary companion to a single star 
\citep[e.g.][]{Santos02,Zucker03}. 
Unrecognized triple systems can be identified indirectly through 
bisector analyses \citep[e.g.][]{Santos02,Eggenberger03} and directly using 
two-dimensional correlation \citep[e.g.][]{Zucker03,Eggenberger07}. 
Due to the strong bias against (moderately) close binaries in the 
samples of regular Doppler planet searches, contamination effects remain 
marginal and do not significantly affect planet searches.


\section{Statistical properties of planetary systems with giant planets}
\label{stat_giants}

The number of planetary companions detected through Doppler spectroscopy 
has increased exponentially since 1995, yielding 
an abundant observational material to characterize the properties of giant 
planets orbiting within 3 AU from their host stars. These data have been 
used extensively to study 
the distributions of planet properties and to quantify the incidence of 
giant planets among different populations of stars 
\citep[e.g.][]{Marcy05a,Udry07,UdrySantos07}. 
As these statistical features are thought to bear fossil imprints of 
planet formation and evolution processes, they provide a crucial 
link between observation and theory. This link has recently been strengthened by 
the new ability of core accretion models to generate synthetic planet 
populations, which can be compared with the observed sample of extrasolar 
planets (\citeauthor{Ida04a}~\citeyear{Ida04a}; 
\citeauthor{Alibert05}~\citeyear{Alibert05}; Mordasini et al. 2009, in prep.).

\subsection{Doppler surveys and their samples}
\label{doppler_surveys}

Radial-velocity planet search programs have traditionally selected their targets to 
optimize the achievable Doppler precision, favoring bright, chromospherically 
inactive main-sequence stars with spectral types between $\sim$F8 and 
$\sim$M0. Among these ``classical'' programs, 
those which have most significantly contributed to the present discoveries 
are: the ELODIE Planet Search \citep[$\sim$330 G-K dwarfs;][]{Perrier03}, 
the California and Carnegie Planet Search 
\citep[$\sim$1100 F-G-K-M dwarfs;][]{Marcy05b}, 
the CORALIE Planet Search \citep[$\sim$1650 G-K dwarfs;][]{Udry00}, 
the Anglo Australian Planet Search \citep[$\sim$230 F-G-K-M stars;][]{Tinney05}, 
and the HARPS Planet Search 
\citep[$\sim$1500 G-K-M dwarfs;][]{Mayor03}. 
To this list we can add the so-called ``metallicity-biased'' surveys, N2K 
\citep[$\sim$2000 F-G-K dwarfs;][]{Fischer05b} and ELODIE\footnote{Now replaced by 
a SOPHIE program \citep{Bouchy07}.} 
\citep[$\sim$1200 G-K dwarfs;][]{daSilva06}, 
which aim at detecting short-period giant planets orbiting 
metal-rich stars. Omitting the metallicity-biased surveys, these programs 
collectively monitor about 3000 F-G-K dwarfs of the solar neighborhood.

Planet searches targeting M dwarfs were originally limited by the 
faintness of their targets \citep{Delfosse98,Marcy98}. Recent instrumental 
developments coupled with the use of larger telescopes have 
considerably improved the situation, allowing the inclusion of a significant 
number of M dwarfs in the present surveys 
\citep{Endl03,Butler04,Bonfils05,Bouchy07}. In total, about 300 
nearby M dwarfs are now being monitored.

As explained in 
Sect.~\ref{limitations}, main-sequence stars earlier than $\sim$F6 are not 
suited for high-precision Doppler measurements and classical surveys avoid them 
accordingly. A recent attempt has been made to extract precise radial 
velocities for A and F dwarfs following a dedicated approach \citep{Galland05}, 
but even with this method planet searches remain limited by stellar activity. 
An alternative approach to study planet occurrence among stars more massive
than the Sun consists in observing these stars when they have evolved off the 
main-sequence, becoming G-K giants or subgiants. Doppler planet searches around 
G-K (sub)giants have gradually expanded during the past five years and some 
hundred stars are now regularly observed 
\citep{Frink02,Setiawan03,Hatzes05,Johnson06,Lovis07a,Niedzielski07,Sato05}. 

A notable built-in bias affecting classical Doppler planet searches is the 
avoidance of binaries closer than 2-6$^{\prime\prime}$ 
(Sect.~\ref{contamination}). Due to this discrimination, current data only 
provide sparse information on the suitability of $\lesssim$200 AU binaries for 
planetary systems. But again the situation is improving, and radial-velocity 
planet searches have recently been extended to spectroscopic and moderately 
close visual binaries \citep[e.g.][]{Eggenberger07,Muterspaugh07}.

Turning to statistical studies, it should be noted that rigorous
analyses including the calculation of detection thresholds and survey
completeness are available only for the oldest programs 
\citep{Cumming99,Endl02,Naef05,Wittenmyer06,Cumming08}, and
with the exception of \citet{Cumming08} these analyses contain too few 
planet detections to study the distributions of planet properties. 
Therefore, 
even though they have not been rigorously corrected for selection effects yet, 
the two surveys best suited for statistical analyses are the CORALIE 
survey with its HARPS extension and the combined Lick$+$Keck$+$AAT survey.

\subsection{Occurrence rate of giant planets}

At present, true occurrence rates are available for the ELODIE and for the Keck 
programs. Results from the ELODIE survey indicate that the fraction of stars 
with a 
planet more massive than $0.5$ M$_{\rm Jup}$ is $0.7\pm0.5$\% for $P < 5$ days 
($a \lesssim 0.06$ AU) and $7.3\pm1.5$\% for $P < 3900$ days ($a \lesssim 4.8$ AU) 
\citep{Naef05}. For the same mass and period ranges, the Keck survey obtains 
$0.65\pm0.40$\% and $8.6\pm1.3$\% \citep{Cumming08}. 

It should be emphasized that according to planet formation models, giant planets 
reside preferentially beyond $\sim$3 AU, while the overall planet population is
dominated (in number) by low-mass planets \citep{Ida04a,Mordasini07}. Taking
into account the selection effects inherent to Doppler spectroscopy, the ELODIE 
and Keck programs could detect only a small fraction of the overall planet 
population ($\sim$6\% according to 
\citeauthor{Mordasini07}~\citeyear{Mordasini07}). 
The above figures thus lend support to the 
hypothesis that planets are common around solar-type stars.

\subsection{Mass, period, and eccentricity distributions}

\subsubsection{Mass distribution}

As explained in Sect.~\ref{kep_analysis}, Doppler spectroscopy does not 
yield the true mass of planetary companions but only the product 
$m_p\sin{i}$, which is a lower limit on the planet mass. 
Nonetheless, as shown by \citet{Jorissen01}, the distribution of minimum 
masses is a good proxy for the mass distribution\footnote{For the sake of
conciseness, we often use the terms ``mass'' and ``mass distribution'' 
instead of ``minimum mass'' and ``minimum mass distribution''. However, we
make the difference or use the additional term ``true mass'' when the 
difference is important.}.

\begin{figure}
\centering
\resizebox{\textwidth}{!}{
\includegraphics{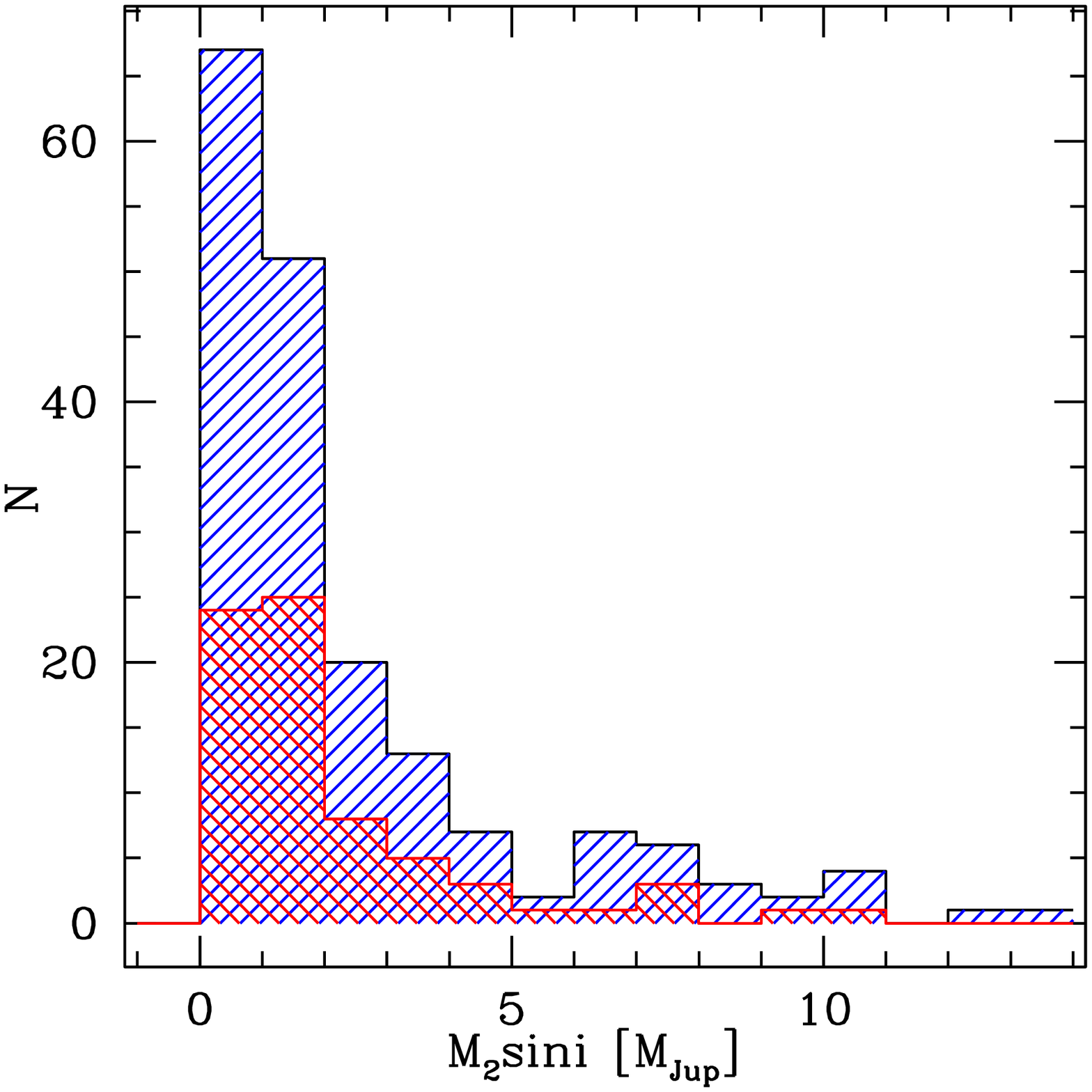}
\includegraphics{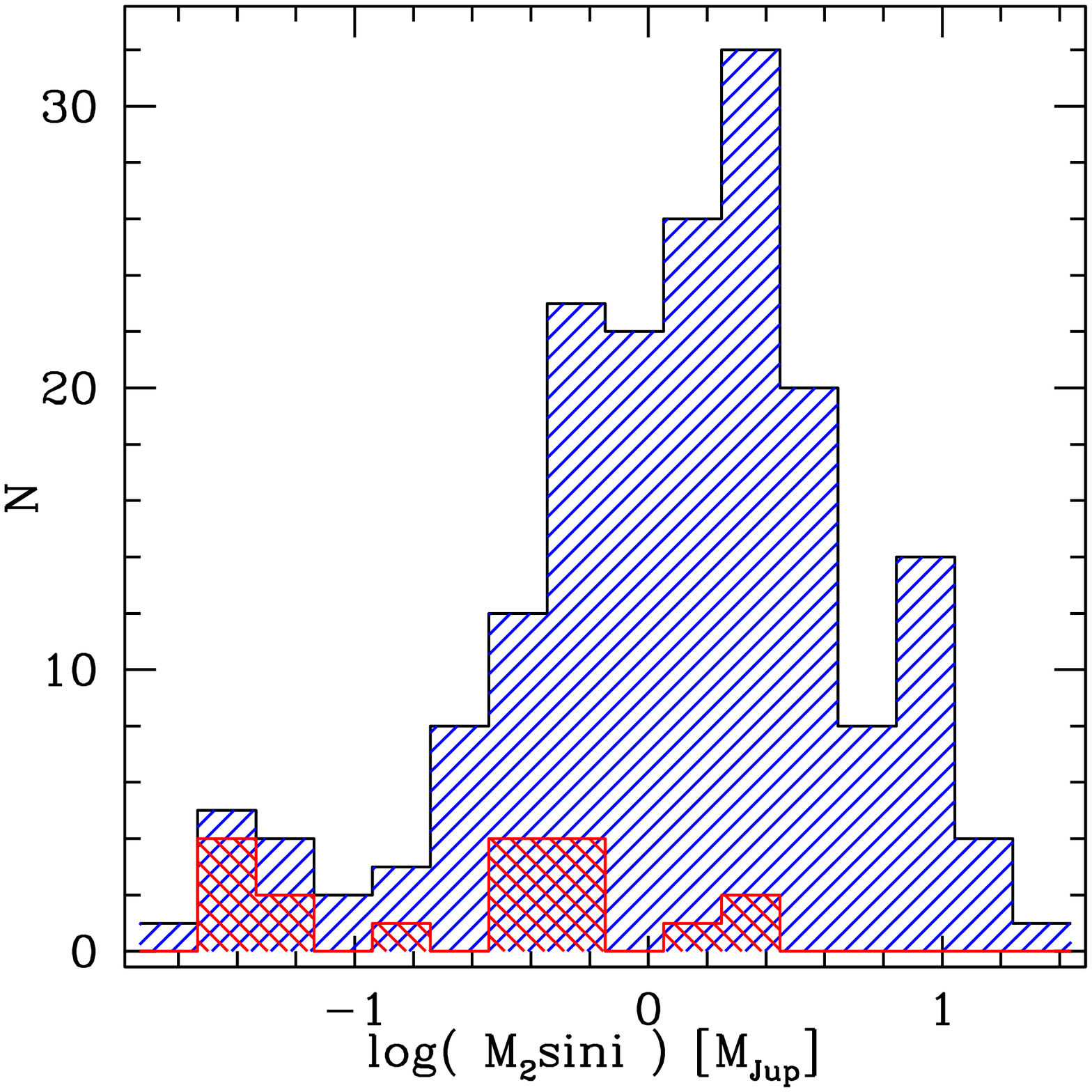}
}
\caption{Distribution of planet minimum masses in linear ({\it left}) and 
logarithmic ({\it right}) scales. The double-hatched histogram represents the 
planets detected with CORALIE ({\it left}) or with HARPS ({\it right}).}
\label{fm}
\end{figure}

The distribution of minimum masses is shown in Fig.~\ref{fm}. 
While most of the detected planets have minimum masses below 5~M$_{\rm Jup}$, 
the distribution exhibits a long tail towards minimum masses larger than 
10~M$_{\rm Jup}$. For orbital periods less than a decade, there is a scarcity 
of companions with (minimum) masses between $\sim$10 and $\sim$100~M$_{\rm Jup}$ 
\citep[e.g.][]{Zucker01}. This gap in the mass distribution 
-- called the brown dwarf desert -- constitutes the most obvious evidence that 
extrasolar planets and stellar companions belong to two distinct populations 
which formed through two different mechanisms. The two populations are not 
completely disconnected, however, as some overlap seems to exist between $\sim$10 and 
$\sim$20~M$_{\rm Jup}$. Distinguishing massive planets from low-mass brown 
dwarfs on the sole basis of their (minimum) mass may thus be hazardous, 
preventing at present a determination of the upper limit to the mass of a
planet.

The low-mass side of the minimum mass distribution is strongly affected by selection 
effects below the mass of Saturn (0.3~M$_{\rm Jup}$), but despite this 
incompleteness the distribution seems to turn upwards near 0.1~M$_{\rm Jup}$. 
This low-mass part of the distribution will be discussed in Sect.~\ref{fm_light}.

\subsubsection{Period distribution}

\begin{figure}
\centering
\resizebox{0.6\textwidth}{!}{
\includegraphics{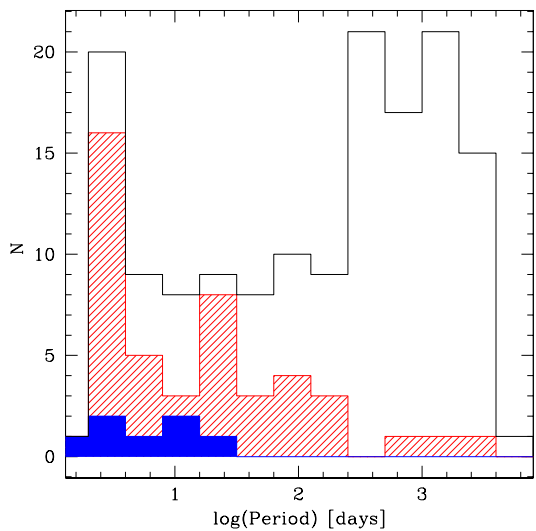}
}
\caption{Period distribution for extrasolar planets. The hatched 
histogram represents light planets with $m_p\sin{i} \leq 0.75$~M$_{\rm Jup}$,
while the filled histogram represents Neptune-mass planets with 
$m_p\sin{i} \leq 21$~M$_{\oplus}$.}
\label{fp}
\end{figure}

The period distribution is shown in Fig.~\ref{fp}. At 
short periods, there is a pile-up of planets with periods between 2 and 5~days. 
These are the so-called hot Jupiters like 51\,Peg\,b. 
The detection of Jovian planets orbiting their stars in just a few days has
been a major surprise since prior to this discovery theory predicted that giant 
planets would form -- and by implication exist -- only beyond a few AU. Although not 
completely excluded, forming these planets in situ is difficult 
\citep{Bodenheimer00} and the first detections of hot Jupiters have 
posed a serious challenge to planet formation theories. According to the revised 
models, short-period giant planets likely formed beyond a few AU and then 
migrated inwards as a result of their tidal interaction with the gaseous disk
\citep[e.g.][]{Lin96,Ida04a,Mordasini07,Armitage07}. The discovery of
close-in giant planets like 51\,Peg\,b has thus revived the notion of 
planetary migration (see the contributions by Terquem and Ida) and there is now 
little doubt that many extrasolar planets have formed at locations which do not 
correspond to those where they are observed today. 

The origin and even the existence of the pile-up of planetary companions with 
periods close to 3 days are still being debated. 
If hot Jupiters have undergone extensive migration within a gaseous disk, 
there is a priori little reason why they should have stopped at small
radii rather than plunge into their star. Some mechanisms have
thus been proposed to stop orbital migration and to park giant planets on
close orbits: the existence of a central 
cavity in the disk \citep[e.g.][]{Lin96,Kuchner02}, tidal interactions with the 
spinning host star \citep[e.g.][]{Trilling98,Paetzold02}, or Roche lobe 
overflow \citep[e.g.][]{Trilling98,Gu03}. Alternative scenarios for the 
formation of hot Jupiters involve resonant interactions with a disk of 
planetesimals \citep{Murray98}, or tidal circularization of highly eccentric 
orbits \citep[e.g.][]{Ford07,Fabrycky07}. 

At longer periods, the number of planets remains fairly constant between 
$\sim$7 and $\sim$250~days, and then increases abruptly. The time baseline of 
ongoing programs is still too short to tell whether the period distribution 
(in logarithmic scale) remains flat from $\sim$300 days to $\sim$4 years, or 
increases continuously beyond $\sim$250~days. 
In either case, current data indicate that there exists a large reservoir of 
giant planets with periods above 5 years ($a \gtrsim 3$~AU) which we are just 
beginning to detect. This shows that the main limitation at the moment for the 
detection of analogs of our Solar System is not the Doppler precision but the 
duration of the surveys.

\subsubsection{Period-eccentricity diagram}

\begin{figure}
\centering
\resizebox{0.7\textwidth}{!}{
\includegraphics{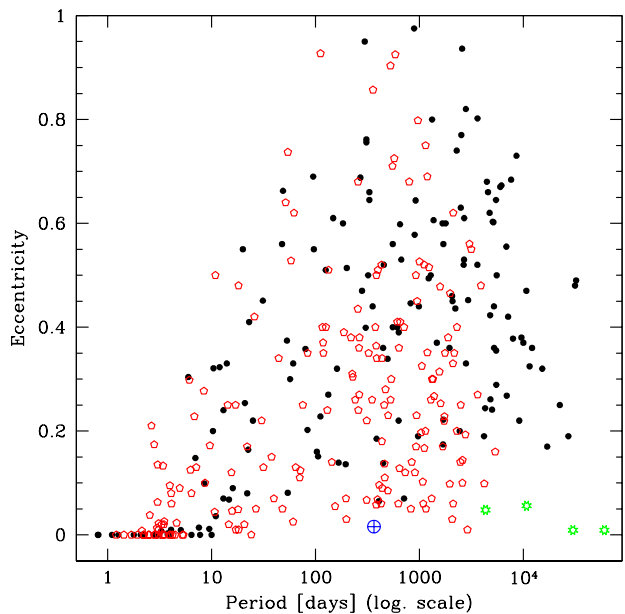}
}
\caption{Period-eccentricity diagram for extrasolar planets (open
pentagons) and for stellar companions to solar-type stars 
(black dots). The Earth and the
four giant planets of the Solar System are also indicated (Earth and starry 
symbols at the bottom).}
\label{pe}
\end{figure}

Figure~\ref{pe} shows the period-eccentricity diagram for both 
extrasolar planets and stellar binaries with solar-type primaries. This plot
makes it clear that most of the extrasolar planets have anomalous 
eccentricities compared with those of Jupiter and Saturn (0.048 and 0.056, 
respectively). Planets with periods below $\sim$5~days tend to have small 
eccentricities due to tidal dissipation \citep[e.g.][]{Rasio96}. Tidal circularization
becomes ineffective at longer periods and planetary eccentricities then span 
almost all the allowable range with a median value of 0.29.

These high planetary eccentricities are very puzzling since formation in a disk 
is thought to yield almost circular orbits. If it is so indeed, some additional
processes must drive the planets out of their initially circular orbits. Numerous  
mechanisms have been proposed to excite planetary eccentricities (see 
\citeauthor{Ford07}~\citeyear{Ford07} for a review and references), including 
strong gravitational scattering between planets, secular interactions with a 
distant companion in a highly inclined orbit, or interactions of orbital 
migration with mean motion resonances (see also Sect.~\ref{multiplanets}). 
While these three mechanisms are all supported by some 
observational evidence, none can reproduce the overall eccentricity 
distribution. It is thus likely that several 
mechanisms contribute in determining the eccentricity distribution. 

Contrary to the mass distribution, the eccentricity distribution of extrasolar 
planets is not radically different from that of stellar companions to 
solar-type stars, though small differences exist \citep[e.g.][]{Halbwachs05}. 
This observation has sometimes been considered as an indication that 
extrasolar planets and stellar companions belong to the same population and 
share a common origin \citep{Stepinski00}. Actually, it is more probable that 
the formation mechanisms were distinct, but that the orbital eccentricities were
excited through similar mechanisms.

\subsection{Multiple planet systems}
\label{multiplanets}

Among the 253 planet-host stars with Doppler measurements, 30 are orbited by 
multiple planets, the most prolific systems being presently 55\,Cnc 
with 5 planets \citep{Fischer08} and $\mu$\,Ara with 4 planets 
\citep{Pepe07}. The observed fraction of multiple planet systems is  
12\%, meaning that the probability of finding a second planet is 
enhanced by a factor of about two with respect to the probability of 
finding the first planet. This multiplicity rate is certainly a lower limit as 
multiple planet systems require more Doppler observations spanning a longer 
time baseline to extract the different Keplerian signals. In the surveys with the
longest duration the fraction of multiple systems ranges from 25\% to 50\%
\citep{FischerValenti05,UdrySantos07,Wright07}. These results suggest that like the Sun, most 
solar-type stars are orbited by systems of planets. 

The presence of two (or several) planets interacting one with another is 
interesting because it increases our potential ability to characterize the 
system's orbital configuration and formation history. For instance, strong
mutual interactions between two planets allow to (partially) remove the 
$\sin{i}$ degeneracy, yielding true planetary masses 
\citep[e.g.][]{Chiang01,Rivera05}. 
From a more specific point of view, the existence of pairs of planets 
locked in low-order mean motion resonances (i.e. with commensurable periods 
such as GJ\,876 b and c) indicates that some planets experienced smooth 
convergent migration \citep[e.g.][]{Lee04,Kley05}. 
Alternatively, the current orbital configuration of the three giant planets
orbiting Ups And suggests that planet-planet scattering occurred in this system 
\citep{Ford05}.

\subsection{Properties of the host stars}

Planet formation being a by-product of star formation, some stellar 
characteristics may provide important information about the conditions leading
to planet formation. Recent work has shown that the
occurrence rate and the properties of giant planets depend on the
mass, metallicity, and multiplicity status of the parent star.

\subsubsection{Stellar metallicity}
\label{metallicity_giants}

\begin{figure}
\centering
\resizebox{0.8\textwidth}{!}{
\includegraphics{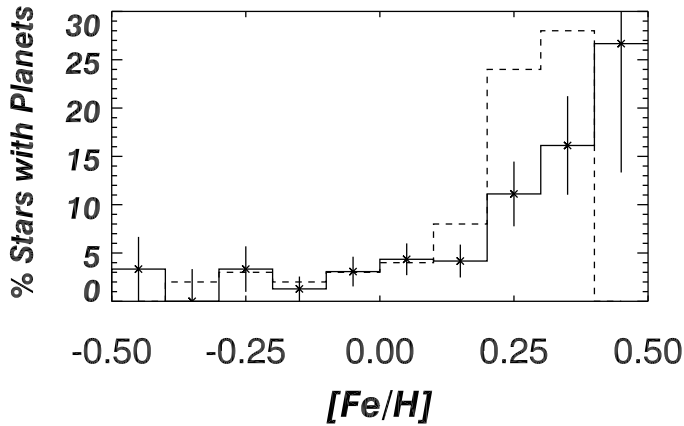}
}
\caption{Percentage of planet-host stars as a function of stellar metallicity.
The dashed line shows the results from \citet{Santos04a} and the solid line shows
the results from \citet{FischerValenti05}.}
\label{probz}
\end{figure}

Not long after the discovery of the first extrasolar planets, it was noticed 
that planet-bearing stars were systematically metal-rich compared to 
the Sun \citep[e.g.][]{Gonzalez97,Santos00b}. Recent studies have shown 
that planet occurrence is a rising function of the parent star metallicity 
(Fig.~\ref{probz}). Specifically, $\sim$25\% of the stars with twice the metal 
content of the Sun ([Fe/H]$= 0.3$) harbor a giant planet, against only 
$\sim$3\% for solar-metallicity stars \citep{Santos04a,FischerValenti05}. This 
observation has motivated the metallicity-biased surveys mentioned in
Sect.~\ref{doppler_surveys}. 

Three explanations have been proposed to understand this correlation: (1) an
enhanced metallicity in the protoplanetary disk could be more conductive to 
planet formation, (2) the infall of metal-rich planetary material onto
the host stars could have enriched their outer layers, or (3) there might be an 
``orbital period bias'' due to the dependence of migration rates on 
metallicity \citep{Livio03} or to the inward shift of the optimum region for 
giant planet formation in metal-poor stars \citep{Pinotti05}. 
Some observational evidence for the last two hypotheses have 
been reported in the literature \citep[e.g.][]{Israelian01,Sozzetti04}, but 
most analyses suggest that the overall metallicity excess observed in 
planet-host stars has a primordial origin and reflects the high metallicity in
the protoplanetary disk 
\citep[e.g.][]{Pinsonneault01,Livio03,Santos03,FischerValenti05}. This
conclusion is corroborated by core accretion models, which reproduce the
observed correlation \citep{Ida04b,Benz06}.   
Disk instability being apparently insensitive to metallicity \citep{Boss02}, 
the metal-rich nature of planet-host stars is commonly considered as a major
evidence that most of the giant planets we observe today formed through core
accretion.

\subsubsection{Stellar mass}
\label{mass_giants}

The occurrence rate of Jovian planets appears to correlate with stellar mass, 
giant planets being rarer around M dwarfs \citep{Endl06,Butler06,Bonfils07} and 
more frequent around intermediate-mass stars \citep{Lovis07a,Johnson07}. 
Another quantity which correlates with stellar mass is the ``average mass of 
planetary systems'' as defined by \citet{Lovis07a}. This second trend may simply 
be a consequence of the first correlation, but it could also indicate that 
planetary systems tend to be more massive around more massive stars. Under the
assumption that the mass of protoplanetary disks scales with stellar mass --
which remains to be confirmed -- both trends are supported by the core accretion 
model \citep{Laughlin04,Ida05,Kennedy08,Benz08}. 

Remarkably, G-K (sub)giants hosting giant planets seem to lack the strong 
metal-rich tendency seen in F-G-K dwarfs 
\citep[e.g.][]{Pasquini07,Takeda08}. 
It is thus likely that both metallicity and stellar mass impact the occurrence 
and the properties of giant planets. Larger samples of planets 
orbiting low- and 
intermediate-mass stars will be needed to confirm this observation and to
disentangle the two effects.

\subsubsection{Stellar multiplicity}
\label{pib}

Contrary to some expectations and despite the strong bias against their
detection, giant planets have been found in various types of binaries and 
triple systems, including spectroscopic binaries with projected
separations of $\sim$20 AU \citep[e.g.][]{Eggenberger04,Raghavan06,Desidera07}. 
Interestingly, some of these planets seem to possess distinctive 
characteristics when compared to planets orbiting single stars: 
(1) the most massive short-period planets are all 
found in binaries \citep[][]{Udry02,Zucker02,Eggenberger04},  
(2) the minimum period for a significant eccentricity seems 
larger for planets in binaries ($\sim$40 days) than for planets around 
single stars \citep[$\sim$5 days;][]{Eggenberger04}, and (3) the four planets
with an eccentricity larger than 0.8 all have a stellar or brown dwarf companion
\citep{Tamuz08}. These trends may indicate that migration and postformation 
evolution proceed differently in some types of binaries than around single 
stars, as suggested by some theoretical studies 
\citep{Kley00,Wu03,Fabrycky07}.


\section{The road to Earth twins}
\label{light_planets}

One of the main drivers of extrasolar planet searches is to understand the 
origin and the degree of uniqueness of a life-bearing planet like the Earth. 
There are thus strong motivations to search for rocky planets orbiting within 
the habitable zones of their stars, that is within the circumstellar 
region where a terrestrial planet can hold liquid water on its surface 
\citep[e.g.][]{Kasting93}. According to theoretical models, such planets  
should be numerous \citep[e.g.][]{Ida04a,Mordasini07}. 
The main obstacle to their discovery through Doppler 
spectroscopy is their low mass, which translates into an extremely 
low-amplitude radial-velocity signal (9 cm\,s$^{-1}$ for the Earth). 

The detection of low-mass planets being closely related to the ultimate 
precision achieved by Doppler surveys, the first Neptune-mass planets 
($10\ \mathrm{M}_{\oplus} < m_p\sin{i} \leq 25\ \mathrm{M}_{\oplus}$) 
were found only four years ago \citep{Butler04,McArthur04,Santos04b}. The
first observation of a so-called super-Earth 
($2\ \mathrm{M}_{\oplus} \leq m_p\sin{i} \leq 10\ \mathrm{M}_{\oplus}$) 
was announced one year later \citep{Rivera05} and opened the way to a series of 
discoveries. 
This new step forward in planet searches was made possible first by 
the development of a new generation of very precise instruments whose prototype is 
the HARPS spectrograph, and second by the application of a careful observing 
strategy to improve detection capabilities and minimize the impact of stellar 
noise.

\subsection{HARPS, a new-generation instrument to search for low-mass planets}
\label{harps}

In 1998 the European Southern Observatory (ESO) issued an announcement of 
opportunity
for the procurement of an instrument dedicated to the search for extrasolar
planet at the unequaled precision of 1 m\,s$^{-1}$. This led to the development
of the HARPS spectrograph, which was manufactured by the HARPS  
Consortium\footnote{The HARPS Consortium is composed of the Geneva Observatory 
(leading institute), the Observatoire de Haute-Provence, ESO, the 
Physikalisches Institut der Universit\"at Bern, and the Service d'A\'eronomie du 
CNRS.}. HARPS is mounted on the ESO 3.6-m telescope at La Silla Observatory 
(Chile) and it has been available to the astronomical community since October 
2003

HARPS (High Accuracy Radial velocity Planet Searcher) is a fiber-fed, 
cross-dispersed echelle spectrograph covering the 
spectral range from 380 to 690 nm with a resolution $R = 115$,$000$ 
\citep{Pepe02,Mayor03,Rupprecht04}. 
Like its two predecessors, ELODIE \citep{Baranne96} and 
CORALIE \citep{Queloz00}, HARPS uses the simultaneous thorium technique. 
But unlike its two predecessors, HARPS is placed inside a vacuum 
vessel, the operating pressure being kept below 0.01 mbar. The vacuum vessel
is in turn located in a thermally stabilized enclosure and this
two-stage insulation ensures a short-term (1 night) temperature 
stability of a few millidegrees. Thanks to this high environmental stability, 
instrumental drifts never exceed 1 m\,s$^{-1}$ during the night, making 
HARPS the most stable spectrograph in the world. 

In exchange for the delivery of the instrument to ESO, the HARPS Consortium 
received Guaranteed Time Observations (GTO; 500 observing nights over 5
years). Since July 2003, the consortium has used this time to carry 
out a comprehensive planet search program \citep{Mayor03}. More than 60\% of 
the GTO time is devoted to the search for low-mass planets around 
$\sim$520 G-K-M dwarfs.

\subsection{Observational results on low-mass planets}

The present sample of low-mass planets (defined here as planets with 
$m_p\sin{i} \leq 25\ \mathrm{M}_{\oplus}$) comprises 
18 objects (\citeauthor{Mayor08b}~\citeyear{Mayor08b} and references therein;
\citeauthor{Forveille08}~\citeyear{Forveille08}). We present below some
noteworthy  
detections made with HARPS in the context of the GTO program. We then 
discuss a few interesting statistical trends that are emerging among this new 
population of low-mass planets.

\subsubsection{Noteworthy HARPS discoveries}
\label{harps_discoveries}

\paragraph{A Neptune-mass planet around $\mu$\,Arae}

The 10.5-M$_{\oplus}$ planet around $\mu$ Arae ($\mu$\,Arae\,c with a period 
of 9.6 days) was found 
serendipitously during an asteroseismology campaign \citep{Santos04b,Bouchy05a}. 
Prior to this detection, 
$\mu$ Arae was known to harbor a giant planet in a 743-day orbit 
\citep{Butler01} and a possible additional planetary companion in a wider orbit
\citep{Jones02}. The continuous monitoring of $\mu$ Arae within the
HARPS GTO program led to the discovery of a fourth planet \citep{Pepe07}, 
making this star host to a rich and complex planetary system. 

\begin{figure}
\centering
\resizebox{0.7\textwidth}{!}{
\includegraphics{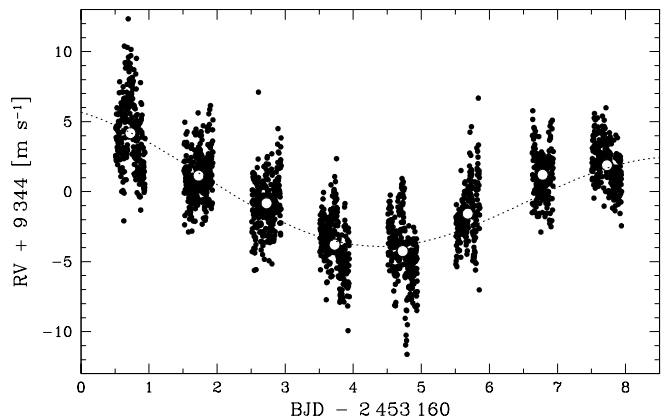}
}
\caption{Radial velocity measurements of $\mu$ Arae obtained during the
asteroseismology campaign. White circles correspond to night averages. The 
dashed line shows the Keplerian orbit of the Neptune-mass planet
confirmed by additional Doppler measurements. Figure from \citet{Bouchy05a}.}
\label{mu_arae}
\end{figure}

As shown on Fig.~\ref{mu_arae}, the amplitude of the radial-velocity signal of 
$\mu$\,Arae\,c is similar to the peak-to-peak radial-velocity variation induced 
by solar-like oscillations. 
The detection of this tiny planetary signal thus strongly benefitted from the 
high measurement density of the asteroseismology campaign. Nonetheless, 
follow-up observations have shown that adopting an adequate observational 
strategy (Sect.~\ref{harps_limitations}), planets such as $\mu$\,Arae\,c are 
within the detection capabilities of the HARPS GTO program.

The quasi-simultaneous discovery of Neptune-mass planets around 55\,Cnc, 
GJ\,436 and $\mu$\,Arae raised the question of their origin and
constitution. Are they sub- or near-critical planetary cores that failed to 
accrete large amounts of nebular gas (see Sect.~\ref{intro}), or are they 
the remnant of gaseous giant planets which lost a large fraction of their 
atmosphere through evaporation? While \citet{Santos04b} favor an essentially 
rocky composition for $\mu$\,Arae\,c, the gas giant remnant and the ice-rock 
composition cannot be excluded \citep{Baraffe06,Brunini05}.

\paragraph{A trio of Neptunes around HD\,69830}

\begin{figure}
\centering
\resizebox{\textwidth}{!}{
\includegraphics[viewport=137 290 564 720,clip]{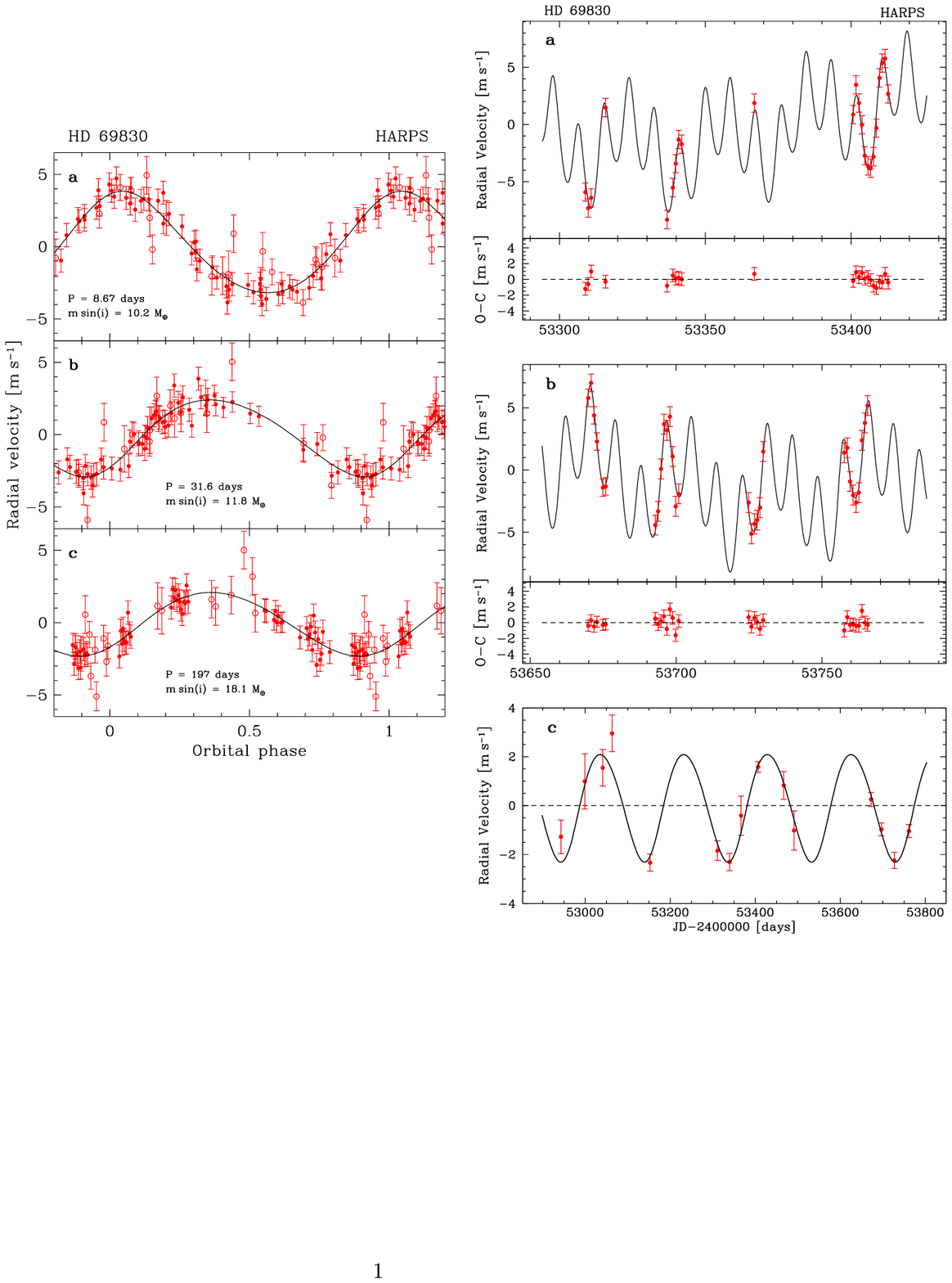}
}
\caption{{\it Left:} Phase-folded orbits for the three planets around 
HD\,69830. {\it Right:} Radial-velocity curve as a function of time. 
Panels a and b show the cumulative signal of the three planets. Panel c shows
the orbit of the outer planet when the observations are binned to one data point
per observing run. Figures from \citet{Lovis06}.}
\label{hd69830}
\end{figure}

Two years ago, \citet{Lovis06} reported the discovery of a planetary system 
made of three Neptune-mass objects (projected masses of 10.2, 11.8, 
and 18.1 M$_{\oplus}$; periods of 8.7, 31.6, and 197 days) 
around HD\,69830 (Fig.~\ref{hd69830}). This system possess two interesting 
characteristics: it is composed exclusively of low-mass planets, and it 
seems to harbor a massive asteroid belt within 1 AU \citep{Beichman05}.  
According to \citet{Alibert06}, the innermost planet is essentially a rocky 
core, while the two outermost planets are rocky 
cores surrounded by a shell of fluid water and a gaseous envelope. 

From the technical perspective, the values of the radial-velocity 
semiamplitudes (2.2, 2.7 and 3.5 m\,s$^{-1}$) make the three planetary signals 
undetectable with most of today's spectrographs. Considering only the most 
recent measurements made with an improved observing strategy, the
root-mean-square (rms) of the residuals around the best-fit solution is 
0.64 m\,s$^{-1}$ and this value decreases to 0.35 m\,s$^{-1}$ 
(bottom right panel of Fig.~\ref{hd69830}) when the 
observations are binned to yield one measurement per observing run 
(typical run duration of 10 days).

\paragraph{Two habitable super-Earths around Gl\,581?}

The M4 dwarf Gl\,581 also harbors a very interesting planetary system made of a 
hot Neptune \citep[$m_p\sin{i} = 16\ \rm{M}_{\oplus}$, period of 5.4 days;][]{Bonfils05} 
and two super-Earths 
\citep[$m_p\sin{i} = 5$ and $8\ \rm{M}_{\oplus}$, periods of 13 and 83 days;][]{Udry07}. 
These two super-Earths lie close to the inner and outer edges of the habitable 
zone of Gl\,581, raising the question of their
habitability. Models of planetary atmospheres indicate that planet c is 
probably too close to Gl\,581 to be habitable \citep{Selsis07a,vonBloh07}. 
Habitability prospects are much better for planet d, which may be an interesting 
target for future space missions aimed at searching for life signatures in the 
atmospheres of low-mass planets.

This system illustrates two advantages M dwarfs possess over solar-type
stars: (1) the smaller mass of M dwarfs makes the detection of low-mass 
planets easier (Eq.~\ref{eq4}), and (2) the habitable zone occurs farther in 
around M dwarfs than around solar-type stars \citep{Kasting93}. These differences imply 
that detecting Earth-mass planets in the habitable zones of M dwarfs is much 
less demanding in terms of Doppler precision than detecting 
the same planets around solar-type stars. The planetary system orbiting 
Gl\,581 shows that detecting potentially habitable planets around 
M dwarfs is within the reach of current high-precision Doppler surveys.

\paragraph{A trio of super-Earths around HD\,40307}

The latest highlight from the HARPS GTO program is the planetary system
discovered around the K dwarf HD\,40307 \citep{Mayor08a}. This system comprises
three super-Earths (minimum masses of 4.2, 6.8 and 9.2 M$_{\oplus}$; periods of
4.3, 9.6 and 20.5 days), the inner planet being the lightest planet detected to 
date around a Sun-like star. Again, one should note the very small 
semiamplitudes of the three reflex motions (2.0, 2.5, and 4.6 m\,s$^{-1}$) and
the low rms of the residuals around the best-fit model (0.85 m\,s$^{-1}$).

\subsubsection{Nascent statistics and emerging trends}

\paragraph{Occurrence rate of low-mass planets}

The recent discoveries of planetary systems with Neptune-mass planets or 
super-Earths confirm that low-mass planets are common in short-period orbits 
around solar-type stars. A first estimate based on the HARPS 
survey suggests that $30\pm10$\%  of the G-K dwarfs harbor a planet with 
$m_p\sin{i} < 30\ \rm{M}_{\oplus}$  and $P < 50$ days \citep{Mayor08a}.

\paragraph{Multiple planet systems with low-mass planets}

The fraction of multiple planet systems with 
at least one low-mass planet is very high ($\sim$80\%). One can argue
that this high value stems from an observational bias, the detection of some 
low-mass planets resulting from a special interest for the host star and its 
first-found planet(s). However, this is less and less true, and many of the recent 
detections are the result of systematic surveys, not of follow-up observations. 
The fraction of multiple planet systems with at least one low-mass planet is 
thus probably intrinsically high.

\paragraph{Mass distribution}
\label{fm_light}

As shown in Fig.~\ref{fm}, the low-mass planets discovered recently seem to 
build up a new population which distinguishes itself from the population of 
giant planets. That is, the mass distribution looks bimodal and the decrease 
previously observed below 1~M$_{\rm Jup}$ may not be due entirely to selection
effects. Remarkably, the synthesis population models of \citet{Mordasini07} 
predict a planetary initial mass function which possess a local minimum
near $\sim$40~M$_{\oplus}$ (0.13~M$_{\rm Jup}$) and a local maximum for 
Neptune-mass planets (0.05~M$_{\rm Jup}$). Present data on low-mass
planets are thus consistent with theoretical expectations and future discoveries
will put to the test the relative overabundance of Neptune-mass planets 
predicted by theory.

\paragraph{Planet-metallicity correlation}

The correlation observed between the occurrence rate of giant planets and
stellar metallicity seems less pronounced   
for low-mass planets \citep{Udry06,Bonfils07}. If confirmed, this trend 
could indicate that  
most of the short-period Neptune-mass planets are subcritical planetary cores 
rather than the remnants of evaporated giant planets. 
Future results from Doppler surveys and from transit 
observations (Sect.~\ref{transits}) should provide important constraints 
to characterize the nature and the origin of short-period Neptune-mass 
planets.

\paragraph{Stellar mass dependence}

M dwarfs seem to harbor more low-mass than giant planets in short-period orbits 
\citep{Endl08,Bonfils07}, implying that the ratio of Jupiter- to Neptune-mass 
planets may decrease with decreasing stellar mass. The sample of low-mass 
planets is still too small to draw robust conclusions, but again this 
trend is consistent with theoretical predictions \citep{Benz08,Ida05,Laughlin04}. 
The planet population orbiting M dwarfs is thus probably quite different from 
the planet population orbiting solar-type stars.

\subsection{Exploring the limitations on Doppler measurements with HARPS}
\label{harps_limitations}

In the context of the GTO program, significant efforts have been 
put into characterizing the limitations on HARPS measurements with the aim  
of identifying the main obstacles to the detection of Earth twins. The two
factors that mainly limit the instrumental precision of HARPS are the
stability in the illumination of the spectrograph and the wavelength
calibration. To minimize variations in the illumination pattern, a specific 
guiding algorithm has been implemented on the ESO 3.6-m telescope. This 
algorithm controls the position of the photocenter at the 0.05 arcsec level, 
keeping the guiding noise below 30 cm\,s$^{-1}$. As to the wavelength
calibration, it has been improved recently and the global uncertainty
is now of 20 cm\,s$^{-1}$ \citep{Lovis07b}.

Stellar noise is another major limitation on very high precision Doppler measurements. 
As shown in Fig.~\ref{solar_oscillations}, solar-like oscillations are clearly 
detected with HARPS. The strategy adopted to minimize this oscillation noise  
consists in setting the exposure times to 10-15 minutes to average
out the signal over 2-3 oscillation periods. Although time consuming, this 
approach allows to keep the residual noise below $\sim$50 cm\,s$^{-1}$ for 
late-G and K dwarfs. On intermediate and long time scales, Doppler
measurements are affected by stellar granulation and stellar activity, respectively. 
The impact of stellar granulation has not been quantified precisely yet. As to 
stellar activity, present results indicate that 
well-selected quiet G-K dwarfs exhibit a jitter level below 1 m\,s$^{-1}$ 
\citep{Pepe08}. Moreover, binning the observations over time scales comparable 
to the rotational periods of the stars permits to significantly average out stellar 
noise. Proceeding in this way, a precision of 30-40 cm\,s$^{-1}$ has been obtained on 
a few test cases (e.g. HD\,69830). Further work will be needed to quantify the 
ultimate limitation imposed by stellar activity, but a careful selection of 
chromospherically inactive targets probably allows to achieve a precision 
better than 50 cm\,s$^{-1}$ on many stars.

The short-term (1 night) precision of HARPS has been characterized 
by asteroseismology observations of the star $\alpha$\,Cen\,B carried out 
during the commissioning phase. Subtracting photon noise (17 cm\,s$^{-1}$) 
and stellar noise (44 cm\,s$^{-1}$) from the global rms of the measurements 
(51 cm\,s$^{-1}$) leaves 20 cm\,s$^{-1}$ for all other error sources 
(guiding errors, instrumental errors, influence of the atmosphere, etc.). On short 
time scales, the precision is thus mainly limited by photon noise 
(targets from planet search programs are fainter than $\alpha$\,Cen\,B) 
and by the intrinsic stability of the stars.

The long-term precision of HARPS is more difficult to characterize. The 
histogram of the observed radial-velocity dispersions for the stars from the 
high-precision GTO program (G and K dwarfs) presents a mode at 
1.4 m\,s$^{-1}$ \citep{Mayor08b}. Yet, part of the observed scatter is likely 
caused by undetected multiple planet systems with low-amplitude Doppler 
signals and by stellar activity. The long-term performances of HARPS are thus 
probably better illustrated by considering the residuals around the best 
orbital solutions published so far. For several stars  
(e.g. HD\,69830, HD\,40307), 
these residuals are below 1 m\,s$^{-1}$, which indicates that HARPS reaches a
sub-m\,s$^{-1}$ precision on the long term. If reproducible
on a sufficiently large sample of stars, the precision of 35 cm\,s$^{-1}$
obtained for HD\,69830 would show that we are not far from detecting  
Earth-mass planets in the habitable zones of Sun-like stars.

\subsection{Perspectives and future instruments}

The exciting results obtained with HARPS have motivated new studies to push 
down the limits of Doppler spectroscopy to reach the 1 cm\,s$^{-1}$ precision 
level. From the instrumental perspective, the experience gained with HARPS 
indicates that reaching this precision level should be possible, though several issues will have
to be solved \citep[][]{Pepe08}. From the scientific perspective, the main 
limitation on ultra high Doppler precision will probably be stellar noise. But 
again the results obtained with HARPS are encouraging, and applying an adequate 
observing strategy on carefully selected targets offers good prospects of
averaging out stellar noise down to 10-20 cm\,s$^{-1}$ on short and 
intermediate timescales for some
dozens of stars. Future ultra-stable high-resolution 
spectrographs are thus under study. The major European project is
CODEX for the Extremely Large Telescope \citep{Pasquini05}. CODEX aims at reaching a
precision of 1 cm\,s$^{-1}$ over at least 10 years to directly measure
the expansion of the Universe. For planet searches, CODEX would provide an 
unprecedented facility for the radial-velocity follow-up of Earth-twin 
candidates detected through photometric transits (see next section). As an 
intermediate step between HARPS and CODEX, ESO foresees the realization of 
ESPRESSO for the Very Large Telescope \citep{Dodorico07}. ESPRESSO would be a 
second-generation HARPS-type instrument designed to achieve a long-term 
precision of 10 cm\,s$^{-1}$. These characteristics would allow to carry out a 
systematic search for Earth-mass planets.


\section{Transiting planets}
\label{transits}

In the course of their orbit, some extrasolar planets intersect the line of 
sight between the star and the observer, momentarily occulting part of the
starlight. These regular passages in front of the stellar disk are called
transits, while passages behind the stellar disk are called 
occultations\footnote{Transits are sometimes called primary eclipses, while 
occultations are also referred to as secondary eclipses or antitransits.}. 
For a distant observer, the probability to see a transiting 
geometry is low for long-period planets such as Jupiter or the Earth 
(0.1\% and 0.5\%, respectively), but is higher for close-in planets such as 
51\,Peg\,b (10\%). The prospects of detecting extrasolar planets by looking for 
transits have thus considerably improved with the discovery of hot Jupiters. 
To date, 52 extrasolar planets are known to transit their parent stars. 
These planets are very precious because they provide us with unique 
information about their physical properties and atmospheres.

\subsection{Observations of planetary transits and occultations}

\subsubsection{Photometric transits}

\begin{figure}
\centering
\resizebox{\textwidth}{!}{
\includegraphics[width=0.45\textwidth]{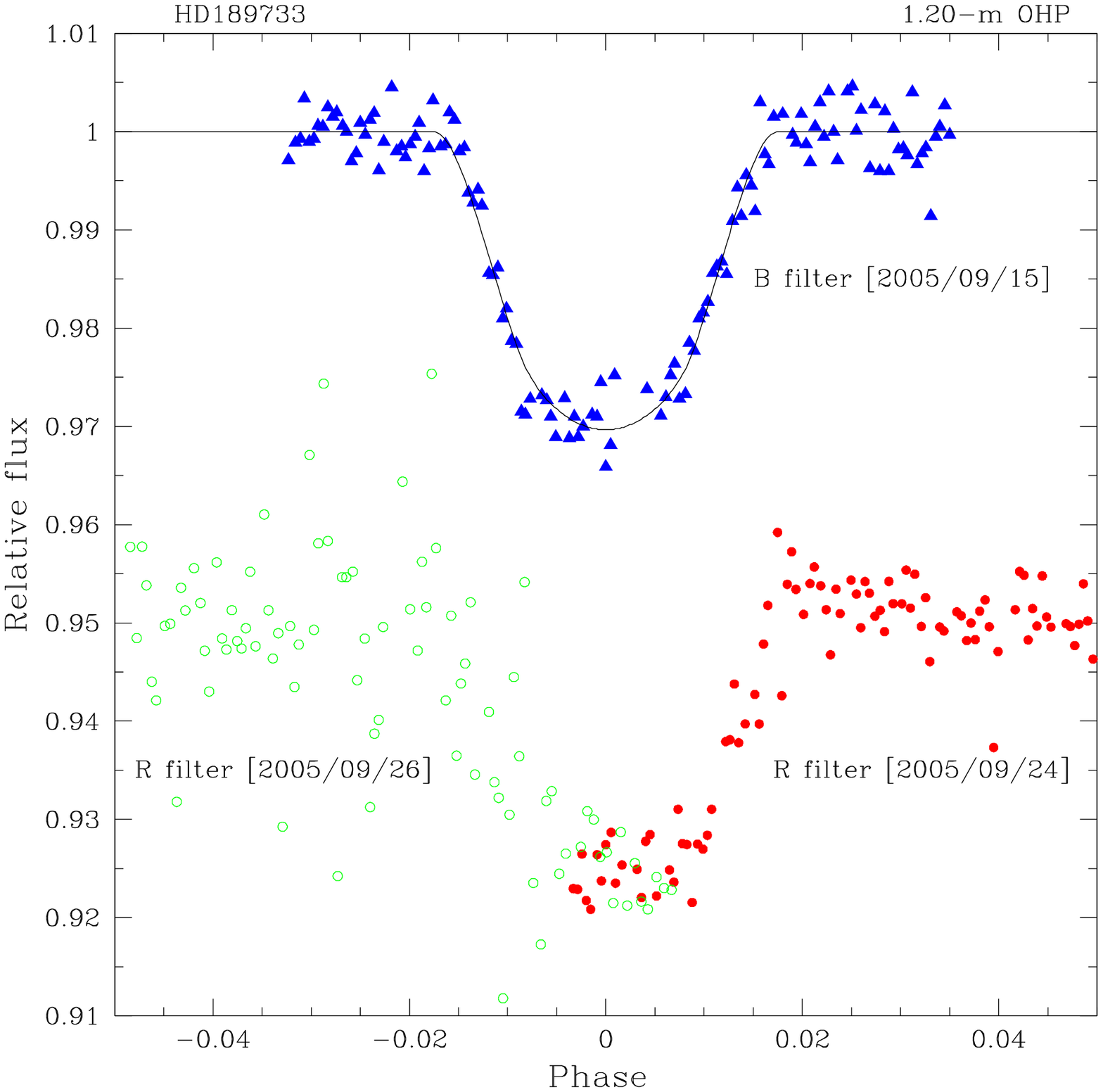}
\includegraphics[width=0.55\textwidth,viewport=15 37 545 468,clip]{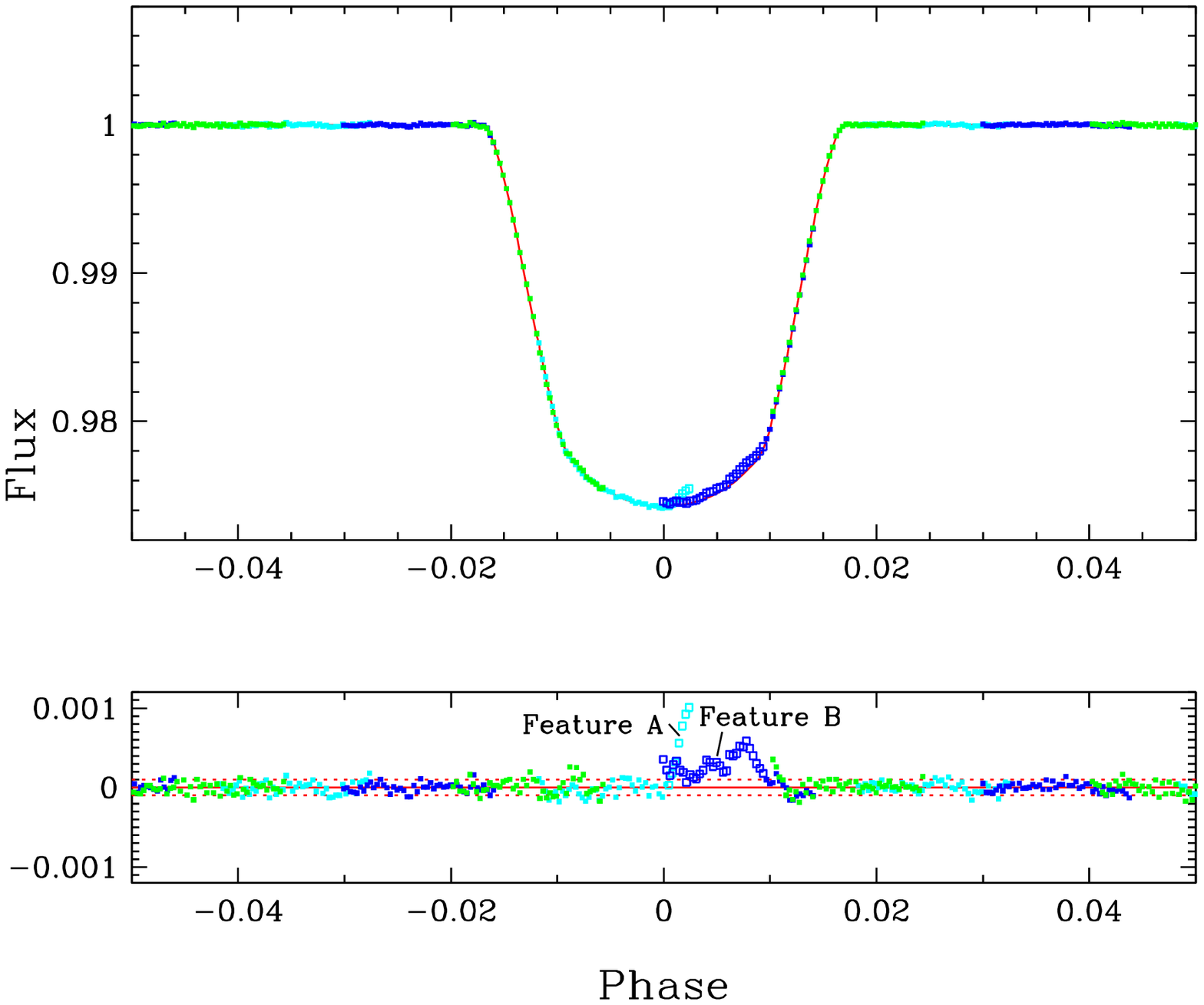}
}
\caption{Phase-folded transit light curves for HD\,189733\,b obtained from the ground 
({\it left}) and from space ({\it right}). Features A and B observed in the
residuals of the right-hand light curve are attributed to the occultation 
of starspots by the planet. Figures from \citet{Bouchy05c} and \citet{Pont07}.}
\label{ex_transits}
\end{figure}

Two examples of transit light curves are shown in Fig.~\ref{ex_transits}. 
Besides their repetition time (which corresponds to the planet's orbital 
period), such light curves provide
four observables: the transit duration, the transit depth, the
ingress/egress duration, and the central curvature. These two last parameters
correspond to relatively subtle features in the curves and can be measured only
on high signal-to-noise data. By supplementing the transit depth, duration, and the orbital
period with some external knowledge of the stellar mass and radius, we can 
estimate the inclination of the planetary orbit and the planet's radius 
\citep[e.g.][]{Seager03}. However, transit photometry provides no information 
about the planet's mass. Transit photometry and Doppler spectroscopy are thus 
highly complementary in their outcomes, and combined observations allow to 
determine both the true mass and the radius of a planet. 

Transiting planets can be discovered and characterized through two different 
paths. In the first case, the planet detection is made through Doppler
measurements and follow-up photometric observations reveal the transit. The
first extrasolar planet found to transit its host star, HD\,209458\,b, was 
discovered in this way \citep{Charbonneau00,Henry00,Mazeh00}. 
In the second case, photometric observations detect a transit and
follow-up Doppler measurements confirm the planetary nature of the 
transiting object. Spectroscopic confirmation is essential in this 
case because transit photometry suffers from a high rate of false alarms 
\citep[e.g.][]{Brown03}. Photometric detection is the most efficient channel to 
discover transiting planets, but the bright and nearby stars of Doppler surveys 
are better suited for complementary studies (Sect.~\ref{fup_studies}). 

The passage of Jupiter in front of the Sun (or of a Neptune-like planet 
in front of an M5 dwarf) produces a dip of 1\% in the light curve, while
a transit of the Earth produces a dip of 0.01\%. 
Photometric transits from 
short-period Jupiter-sized planets can be detected from the ground, but 
transits from Earth-sized planets are detectable only from space (see e.g. 
\citeauthor{Brown08}~\citeyear{Brown08} for further details). 
Numerous photometric surveys searching for planetary transits are presently 
active and the rate of discovery has increased rapidly over the past two years. 
Among ground-based surveys, HAT \citep{Bakos07}, OGLE \citep{Udalski02a}, 
TrES \citep{Alonso04}, WASP \citep{CollierCameron07},
and XO \citep{McCullough06} have successfully detected transiting
planets. These surveys typically monitor tens of thousands of 
stars with a photometric precision better than 10 mmag. 
CoRoT -- the first space-borne mission largely dedicated to the discovery of 
planetary transits -- has also found several planets since its launch in
December 2006 \citep{Barge08}. During its 2.5-year mission, CoRoT will monitor 
$\sim$60,000 stars for 150 continuous days, permitting the detection of
super-Earths ($\geq$2 R$_{\oplus}$) and larger planets with periods 
below $\sim$50 days \citep[e.g.][]{Moutou06b}.

\paragraph{An example of photometric follow-up: GJ\,436}

The low-mass planets discovered since 2004 have prompted intensified 
follow-up photometry to check for possible transits. 
The first -- and currently only -- Neptune-sized planet found to transit its parent 
star is GJ\,436\,b, whose transits were detected at a non-professional observatory 
in the Swiss Alps \citep{Gillon07a}. This discovery opened a new window for 
planetary studies and triggered immediate follow-up observations with
the Spitzer Space Telescope \citep[][see also Sect.~\ref{fup_studies}]{Gillon07b,Deming07,Demory07}. The mass 
and radius of GJ\,436\,b suggest an internal structure similar to that of
Neptune, with a rocky core surrounded by a thick water layer and a thin 
gaseous envelope (Fig.~\ref{mass-radius}). Nonetheless, without additional  
observational constraints, the internal composition of GJ\,436\,b cannot be
determined in detail  
\citep[][see also Sect.~\ref{mradius}]{Adams08}.

\paragraph{An example of Doppler follow-up: the OGLE candidates}

In 2002-2003, OGLE was the first photometric survey to deliver a series of 
planet candidates \citep{Udalski02a,Udalski02b,Udalski02c,Udalski03}, 
and Doppler follow-up programs were immediately initiated to identify genuine
planets. The task was difficult, however, because most of the OGLE targets 
are too faint ($V \in$ [14;18]) to be observed with standard planet-search 
spectrographs. The Geneva team resorted to using the FLAMES/UVES 
multi-fiber spectrograph on the Very Large Telescope (VLT) to follow up 
spectroscopically many of the OGLE candidates \citep{Bouchy05b,Pont05}. This 
effort has been very successful, leading to the discovery  
of 3 of the 5 transiting planets found among the 137  
first OGLE candidates \citep{Bouchy04,Pont04}. The follow-up campaign 
has also yielded empirical occurrence rates for the main types of 
false alarms: transits by small stars (38\%), multiple star systems (24\%), 
false transit detections (14\%), and grazing eclipses (7\%). Planets
represent only $\sim$7\% of the total number of candidates. 

The follow-up of the OGLE candidates has demonstrated the necessity and 
the efficiency of Doppler observations in identifying genuine planets among the
numerous false alarms. Yet, characterizing the transits exhibited by the
faintest OGLE targets is close to being impractical. The OGLE 
survey is thus near the upper limiting magnitude for spectroscopic follow-up  
with existing ground-based facilities.

\begin{figure}
\centering
\resizebox{0.7\textwidth}{!}{
\includegraphics[viewport=1 185 550 550,clip]{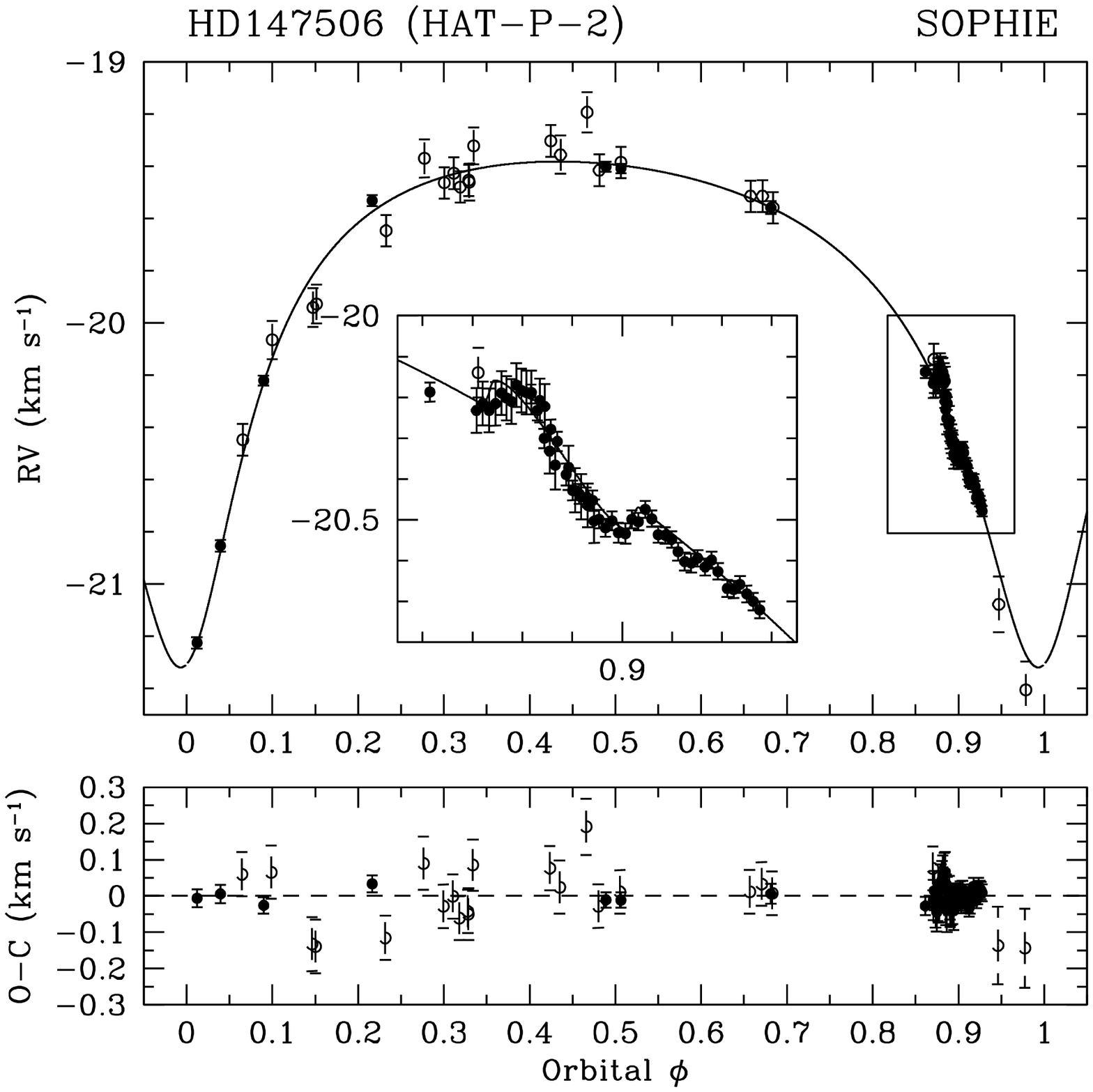}
}
\caption{Phase-folded radial-velocity curve for HD\,147506 showing the
Rossiter-McLaughlin effect (inset). Figure from \citet{Loeillet08}.}
\label{rm}
\end{figure}

\subsubsection{Spectroscopic transits: the Rossiter-McLaughlin effect}

A planetary transit produces not only a photometric signal but also a 
spectroscopic signal called the Rossiter-McLaughlin effect. 
This anomaly in Doppler shift measurements traces the passage of the planet 
in front of the rotating stellar disk (Fig.~\ref{rm}). That 
is, when the planet occults part of the approaching (blueshifted) stellar 
hemisphere, the integrated starlight appears slightly redshifted, an effect 
interpreted as an anomalously high radial velocity. The reverse effect is 
observed when the planet occults part of the receding stellar hemisphere. 
For extrasolar planets, the Rossiter-McLaughlin effect was first observed and 
modeled for HD\,209458 \citep{Queloz00}. Since then it has been detected for 
11 transiting systems (see \citeauthor{Winn08}~\citeyear{Winn08} for a review). 
Two important parameters can be obtained by 
modeling the Rossiter-McLaughlin effect:  
the direction of planetary revolution (prograde or retrograde orbit), and 
the angle $\lambda$ between the sky projections of the orbital and the 
stellar rotation axes. This angle gives us some information about the
degree of alignment of the planet's orbital angular momentum vector with the
stellar spin axis (spin-orbit alignment), which provides in turn an important
constraint for the formation models of hot Jupiters (Sect.~\ref{sp_alignment}).

\subsubsection{Precise monitoring of transits and occultations}
\label{fup_studies}

Close-in planets with large planet-to-star radius ratios and transiting nearby 
bright stars permit a variety of follow-up studies which provide us with 
deeper insight into the physical properties of these planets. 
The two most favorable systems for detailed studies are 
HD\,209458\,b \citep{Henry00,Charbonneau00,Mazeh00} 
and HD\,189733\,b \citep{Bouchy05c}. 

Transit light curves of extreme accuracy -- as obtained with the Hubble Space 
Telescope, e.g. right panel of Fig.~\ref{ex_transits} -- reduce the ambiguity between the estimated parameters (stellar 
radius, planetary radius, orbital inclination, stellar limb darkening) and 
yield very precise planetary radii \citep[e.g.][]{Brown01a,Pont07}. 
High-quality transit time series allow additionally to search for evidence of 
planetary satellites, Saturn-type rings, or additional planets 
\citep[e.g.][]{Sartoretti99,Barnes04,Agol05}.

Follow-up observations using the Hubble\footnote{http://hubble.nasa.gov} and 
Spitzer\footnote{http://www.spitzer.caltech.edu} Space 
Telescopes have provided the first data on the atmospheres of extrasolar 
planets. At visible and ultraviolet wavelengths, the light emitted by the 
planet is negligible and the technique of transmission spectroscopy  has been 
employed to study the chemical composition and the physical properties 
of planetary atmospheres \citep[e.g.][]{Brown01b}. 
Transit spectroscopy uses the fact that the major 
sources of opacity in a planetary atmosphere depend on wavelength. Due to this 
dependence, the planet's effective radius -- and thus the transit depth, which 
is the ratio of the planet-to-star surface area -- 
vary with wavelength. Searching for relative 
changes in transit depth as a function of wavelength is thus a means to probe 
the absorption properties of a planet's atmosphere 
\citep[e.g.][]{Charbonneau02,Vidal-Madjar03}.

At infrared wavelengths, the planet-to-star contrast ratio improves 
considerably, permitting a direct detection of the photons emitted from the 
planet. Since hot Jupiters are believed to rotate synchronously with their 
orbital period (i.e. they are tidally locked), they present permanent day and 
night sides. The depth of the secondary eclipse (occultation) allows to 
characterize the day-side flux (thermal emission) from the planet 
\citep[e.g.][]{Deming05,Charbonneau05}, 
while observations at various orbital phases allow to characterize the 
planet's longitudinal temperature profile \citep[e.g.][]{Harrington06,Knutson07}. 
In addition, the timing and duration of the secondary eclipse place useful 
constraints on the orbital eccentricity \citep[e.g.][]{Deming07,Demory07}. Up to now, the
thermal emission from six extrasolar planets has been detected  
\citep[e.g.][and references therein]{Charbonneau08}.

\subsection{Noteworthy results obtained from transiting planets}

\subsubsection{A new class of planets: the very hot Jupiters}

Three of the five planets discovered among the first 137 OGLE candidates have 
orbital periods of $\sim$1.5 days \citep{Konacki03,Bouchy04}, 
which originally placed them in an empty region 
of the period distribution, below the pile-up of hot Jupiters. These new 
planets were accordingly termed very hot Jupiters. Since the sensitivity of 
Doppler spectroscopy increases with decreasing period, the absence of very 
hot Jupiters in the discoveries of radial-velocity surveys looked surprising. 
\citet{Gaudi05} showed that transit surveys are actually $\sim$6 times more 
sensitive to planets with periods of 1 day than of 3 days, a fact that 
could explain the observations. Since then, Doppler surveys have found  
several planets with periods of 2-3 days, while transit surveys have 
detected many regular hot Jupiters. These discoveries have bridged the gap 
between the two populations and confirm that there is no significant 
incompatibility between the results of Doppler and transit surveys.

\subsubsection{Mass-radius relationships and bulk compositions}
\label{mradius}

\begin{figure}
\centering
\resizebox{0.7\textwidth}{!}{
\includegraphics{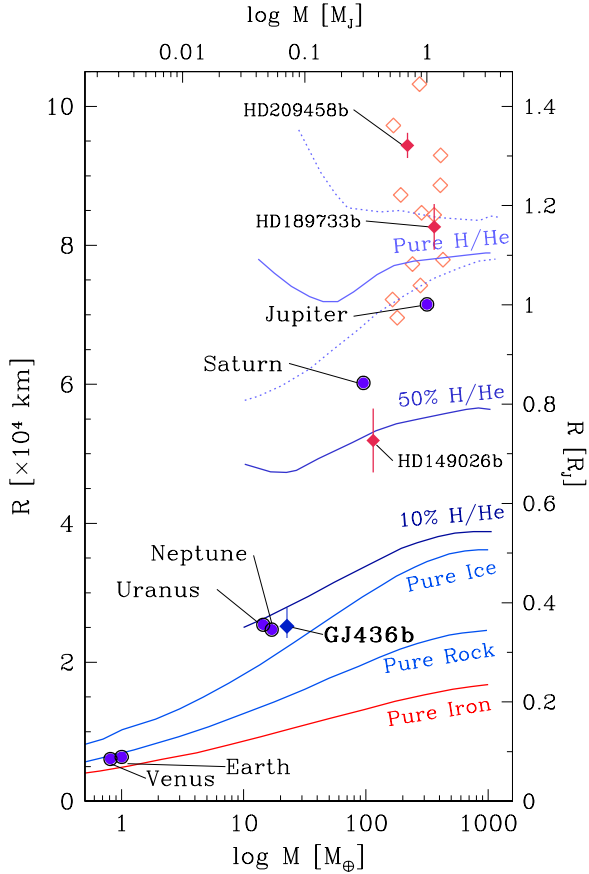}
}
\caption{Mass-radius diagram for planets. Transiting hot Jupiters 
(diamonds), planets from the Solar System (dots), and 
GJ\,436\,b are indicated. Figure from \citet{Gillon07a}, adapted
from the models of \citet{Fortney07}.}
\label{mass-radius}
\end{figure}

The measurement of a planet's mass and radius give us the mean
density, from which we can infer some information about the planet's bulk 
composition (Fig.~\ref{mass-radius}). 
Most of the known transiting planets have densities below 
1.5 g\,cm$^{-3}$, so they must be composed primarily of hydrogen and helium, 
like Jupiter and Saturn. Historically, this has been an important confirmation 
of their planetary status. 

Some transiting giant planets have high densities which cannot be
explained by a pure H/He composition. The most extreme case is 
HD\,149026\,b (indicated on Fig.~\ref{mass-radius}), whose small radius 
requires the presence of $\sim$70 M$_{\oplus}$ of heavy elements 
(2/3 of the total mass) in its interior 
\citep{Sato05}. Giant planets with small radii are accounted
for due to variations in the amount of heavy elements sequestered in a central   
core and/or distributed throughout the gaseous envelope 
\citep[][]{Guillot08,Burrows07c}. Since the presence 
of a solid core is a natural characteristic of formation by core accretion, giant 
planets with small radii appear to support the core accretion model. 
Nonetheless, if they are common, planets with 2/3 of their mass 
of heavy elements are a challenge to any formation model, including core
accretion \citep{Sato05,Ikoma06,Broeg07}. 

Contrary to HD\,149026\,b, some giant planets such as HD\,209458\,b (also 
indicated on Fig.~\ref{mass-radius}) have unexpectedly large radii and low 
densities. According to standard models of irradiated planets, these hot
Jupiters are too big for their mass and age, a characteristic which is still  
unexplained (see \citeauthor{Guillot08}~\citeyear{Guillot08} or 
\citeauthor{Burrows07c}~\citeyear{Burrows07c} for further details). 
 
Low-mass planets are expected to display a wide variety of compositions and
internal structures. According to theoretical predictions 
\citep[e.g.][]{Fortney07,Seager07}, Neptune-mass 
planets can be composed primarily of refractory materials (rock, iron, 
planetary ices\footnote{Planetary ices comprise ammonia, methane, and water 
(in any phase, not necessarily solid).}, 
but also of H/He, or of a mixture of these possibilities. 
As to super-Earths, they are presumably composed of 
refractory species. It should be noted that there is some ambiguity and 
degeneracy in inferring bulk compositions from mean densities, especially for
planets with a mass between 5 and 20 M$_{\oplus}$ \citep{Adams08}. 
For example, mass and radius measurements alone are insufficient to distinguish 
between an ocean planet (a planet with $>$25\% of water by mass) and a rocky 
planet with a massive H/He gaseous envelope.

\subsubsection{Spin-orbit alignment}
\label{sp_alignment}

Rossiter-McLaughlin observations obtained so far show that in all except 
one case transiting planets are on prograde orbits aligned with the stellar 
rotation axis (see
\citeauthor{Winn08}~\citeyear{Winn08} for a review). These
results suggest that the migration of hot Jupiters generally preserves 
spin-orbit alignment, an observation that disfavors scattering processes or 
Kozai migration as the main formation channel for hot Jupiters.

\subsubsection{Planetary atmospheres}
\label{results_atm}

Hot Jupiters are tidally locked and 
receive $\gtrsim$10,000 times more radiation from their stars than Jupiter does from 
the Sun. Due to this high and asymmetric irradiation, their atmospheric properties 
are expected to differ significantly from those of Jupiter and Saturn. 
Atmosphere models of hot Jupiters predict transmission spectra characterized by 
strong absorption lines due to the sodium and potassium resonance doublets at
visible wavelengths, and by strong molecular bands of water, carbon monoxide, 
and methane in the infrared \citep[e.g.][]{Brown01b}. 
A significant source of uncertainty in atmosphere models is the possible 
presence of hazes or clouds, which can block the stellar 
flux at any height in the atmosphere. Another key question regarding hot 
Jupiters is whether atmospheric circulation efficiently redistributes the
absorbed stellar radiation from the day side to the night side.

Follow-up studies on HD\,209458\,b have revealed the presence of sodium, 
hydrogen, water, and more tentatively oxygen and carbon in its atmosphere 
\citep{Charbonneau02,Vidal-Madjar03,Vidal-Madjar04,Barman07,Burrows07b,Knutson08c}. 
Hydrogen absorption extends beyond the planet's Roche lobe\footnote{The
Roche lobe outlines the volume surrounding an object within which material is
gravitationally bound to it.}, 
indicating that HD\,209458\,b has an extended and evaporating atmosphere 
(see the contribution by Ehrenreich). 
The infrared broadband emission spectrum of HD\,209458\,b differs significantly
from the predictions of standard atmosphere models, which was interpreted as
evidence for an atmospheric temperature inversion \citep{Knutson08b,Burrows07b}. 
The day-night contrast of HD\,209458\,b has not been measured directly, but 
atmospheric circulation seems relatively efficient \citep{Cowan07,Knutson08c}.

Follow-up studies on HD\,189733\,b engendered lively discussions about the 
detection of water in its atmosphere
\citep{Tinetti07a,Ehrenreich07,Beaulieu08,Swain08b,Charbonneau08,Barman08,Fortney07b,Grillmair07}.
The good agreement obtained recently between the broadband emission 
spectrum measured by 
\citet{Charbonneau08} and the models of \citet{Barman08} provides convincing 
evidence that water absorption does shape the infrared spectrum of 
HD\,189733\,b. Methane and sodium have also been detected in its atmosphere 
\citep{Swain08b,Redfield08}. The emergent spectrum of HD\,189733\,b is well 
matched by standard models which do not include an atmospheric temperature 
inversion \citep{Charbonneau08}. 
HD\,189733\,b is only slightly hotter on the day side than on the night side, 
indicating that the energy from the irradiated hemisphere is efficiently 
redistributed throughout its atmosphere \citep{Knutson07,Knutson08b}.

Detailed follow-up studies on HD\,209458\,b and HD\,189733\,b have thus revealed
both similarities and fundamental differences in their atmospheric 
properties. Together with the data obtained on a few other transiting 
systems (see \citeauthor{Burrows08}~\citeyear{Burrows08} for a review), the 
above results support the idea that the atmospheres of hot Jupiters diverge 
into two groups: those with and those without a temperature inversion 
\citep{Fortney08,Burrows08}.

\subsection{Perspectives and future instruments}
 
In 2009, CoRoT will be joined by the Kepler\footnote{http://kepler.nasa.gov}
satellite, which will monitor 100,000 
main-sequence stars for 4 continuous years to detect the transits of 
Earth-like planets \citep[e.g.][]{Basri05}. Besides the discovery of hundreds of 
low-mass planets, CoRoT and Kepler will provide the first measure 
of the occurrence of Earth- and Neptune-sized planets at short and intermediate  
periods. The frequency of Earth-sized planets orbiting in the habitable zones of 
solar-type stars is a key parameter for the design of the next generation of 
project missions, which will aim at characterizing the atmospheres of 
Earth-like planets.

Space-based transit searches do not eliminate the need for spectroscopic
follow-up, but the higher quality of the photometry reduces the rate of false
alarms. The radial-velocity follow-up of the CoRoT planet candidates 
is conducted with CORALIE, SOPHIE, HARPS, the Coud\'e spectrograph on the 
Alfred-Jensch telescope, and FLAMES/UVES for the faintest targets. 
The spectroscopic follow-up of the Kepler planet candidates will be much more 
challenging and will require a larger investment of telescope time. 
To perform the Doppler follow-up of interesting planet candidates identified by 
Kepler, the Harvard University and the Geneva Observatory have joined in a 
collaboration to build and operate the HARPS-N 
spectrograph\footnote{http://obswww.unige.ch/Instruments/harps\_north}. 
This improved HARPS-type instrument will be installed 
on the William-Herschel Telescope at La Palma Observatory in 2009. In
synergy with Kepler, HARPS-N is a promising step towards the
detection of rocky planets in the habitable zones of solar-type stars.

Regarding the precise monitoring of transits and occultations, Spitzer will run 
out of cryogen in spring 2009, which will eliminate its ability to 
observe at wavelengths longer than 4.5 $\mu$m. An extended ``warm'' Spitzer 
mission has been proposed, in which the spacecraft would continue to operate 
(at the same sensitivity) in the two shortest wavelength channels. Unlike
Spitzer, Hubble should be reborn with the servicing mission scheduled for early
2009. Beyond Hubble and Spitzer, we will have to wait the launch of the 
James Webb Space Telescope\footnote{http://www.jwst.nasa.gov} in the middle of 
the next decade to continue and extend the study of extrasolar planetary atmospheres.


\section{Conclusion and perspectives}
\label{conclusion}

For centuries, our knowledge of planetary systems has been based on a single 
example, the Solar System. During the last thirteen years, Doppler spectroscopy
has allowed the detection of $\sim$300 extrasolar planets, opening a new window 
on planetary sciences. These discoveries have considerably broadened our 
appreciation of the diversity of possible planetary systems and have 
revolutionized our ideas about planet formation.

By the standards of the Solar System, many extrasolar planets exhibit 
singular properties, including very short orbital periods, high eccentricities, 
and large masses. Stars of different masses seem to harbor different planet 
populations and multiple planet systems appear to be the rule. For solar-type 
stars, stellar metallicity is a strong predictor of the presence of giant 
planets in the inner regions of the system. 
Migration via tidal interaction with the gaseous protoplanetary disk 
appears to play a central role in the formation and early evolution 
of these giant planets. Some extrasolar planets were found in environments 
relatively hostile to their presence (pulsars, spectroscopic binaries), 
indicating that planet formation is a robust process. The recent discoveries of 
Neptune-mass planets and super-Earths at close (Doppler spectroscopy) and 
intermediate (microlensing) separations support the theoretical prediction that 
low-mass planets are more numerous than their giant cousins.

Transiting planets are only a small subset of all extrasolar planetary
systems, but they offer an unique opportunity to deepen our understanding of
planet structure and composition. With the exception of one hot Neptune, 
transiting planets are gas giants with a wide range of masses and radii. 
Follow-up observations using the Hubble and Spitzer Space Telescopes provided 
the first glimpse into the atmospheres of some hot Jupiters, and the first 
detections of photons emitted from extrasolar planets. Planetary emission 
spectra reveal fundamental differences in atmospheric properties, leading to the 
suggestion that the irradiated atmospheres of hot Jupiters bifurcate into two 
groups. 

The above observations provide an ever-growing database with which planet 
formation theories can be put to the test. By combining observed  
properties of protoplanetary disks, models of planet formation and evolution, 
and observational selection effects, theorists now generate 
in a self-consistent
way synthetic planet populations which can be compared with the actual 
detections. Recent models based on the core accretion paradigm successfully 
reproduce key observational features such as the mass distribution and the 
planet-metallicity correlation. Although much remains to be understood about 
planet formation, these new theoretical achievements reinforce 
the idea that most of the known planets formed through core accretion. 

To progress in our understanding of planet formation and to appreciate the 
significance of our Earth, we must detect a wide range of planetary systems, 
including planets and host stars both similar and dissimilar to our own. 
This can be achieved through two
different paths: by further developing existing techniques to increase their 
capabilities, and by developing new techniques to explore different regions of 
the parameter space. The characterization of transiting planets illustrates the 
benefit of combining the results obtained with two different observational 
techniques.

As we have seen, Doppler spectroscopy has reached neither its instrumental
nor its astrophysical limits. The latter will probably be dominant, but 
applying an adequate strategy on carefully selected targets 
offers good prospects of averaging out stellar noise 
down to 10-20 cm\,s$^{-1}$ on short and intermediate time scales for some 
dozens of stars. The next generation of HARPS-type instruments aiming at a 
Doppler precision of 10-20 cm\,s$^{-1}$ (HARPS-N, the ESPRESSO project) should 
therefore be able to detect Earth-mass planets in short-period orbits, and to 
characterize some of these planets up to periods of $\sim$1 year. 
Ultra-stable spectrographs aiming at a Doppler precision of 1 cm\,s$^{-1}$
should be technically feasible. If stellar noise permits, such instruments would
allow the detection and the precise characterization of Earth twins. 

While Doppler spectroscopy is intrinsically sensitive to short-period planets,
astrometry and direct imaging are particularly suited to detect planets in wide
orbits. Up to now, these two techniques have very marginally contributed to 
planetary studies, but this is going to change thanks to the upcoming 
deployment of new facilities. In 2009, the 
PRIMA\footnote{http://obswww.unige.ch/Instruments/PRIMA} instrument for the VLT 
interferometer should become fully operational, allowing the astrometric 
detection of planets down to the mass of Neptune in the separation range 
1-5 AU \citep{Launhardt08}. From space, the 
GAIA\footnote{http://gaia.esa.int/science-e/www/area/index.cfm?fareaid=26} 
mission (scheduled for launch in 2011) will detect thousands of extrasolar 
planets, revolutionizing the database of extrasolar planets with known orbital
elements and masses \citep{Casertano08}. 
Although astrometric observations are sufficient to determine 
three-dimensional orbits and true planetary masses, additional Doppler 
measurements will be of great help in constraining the large number of free 
parameters, especially for multiple planet systems. Regarding direct imaging, 
a few ``extreme'' adaptive optics systems are under construction, including 
SPHERE for the VLT \citep{Beuzit08} and GPI for the Gemini Observatory 
\citep{Macintosh08}. Starting in $\sim$2011, these instruments will detect, 
characterize the atmospheres, and determine the radii of wide ($\sim$5-100 AU) 
and massive ($\gtrsim$1-5 M$_{\rm Jup}$) planets of various ages 
($\sim$10-1000 Myr). All together, these upcoming facilities will 
probe the intermediate and outer regions of protoplanetary disks, providing 
new information about the formation, evolution, and migration of giant planets.

Doppler spectroscopy and microlensing have opened the way to the detection of
low-mass planets, and mature detection techniques now aim at 
identifying an Earth-mass planet located in the habitable zone of its host 
star. Microlensing surveys can yield the frequency of such planets, but they do 
not allow a precise characterization of 
their planet candidates \citep[e.g.][]{Beaulieu08}. Space-based transit 
photometry is presently the most sensitive method for planetary detection, and 
the pending Kepler mission offers good prospects of finding the first 
Earth-sized planet potentially habitable. Yet, Doppler spectroscopy is
a serious competitor, and it may win the race. In any case, both techniques 
will play a key role in detecting and characterizing Earth-mass planets in the 
near future. Due to their ubiquity and low mass, M dwarfs are ideal targets for 
the detection of terrestrial planets. In addition to optical radial velocities, 
infrared Doppler spectroscopy represents an attractive new technique to search 
for and characterize planets around late-type ($\gtrsim$M4) M dwarfs 
(e.g. \citeauthor{Martin05}~\citeyear{Martin05} or the SPIRou 
project\footnote{http://www.ast.obs-mip.fr/article.php3?id\_article=637}). 
In the more distant future, space-borne astrometry could contribute to the 
detection of low-mass planets as well. 
The SIM Lite\footnote{http://planetquest.jpl.nasa.gov/SIMLite/sim\_index.cfm} project 
would serve this purpose \citep[e.g.][]{Tanner06}, but its future is highly uncertain. 

If it is approved, the warm Spizter mission will allow the community to retain 
the ability to 
characterize the atmospheres of some transiting planets for a few more
years. In the middle of the next decade, JWST will take over and will allow  
to study the atmospheres of Earth-sized planets located in the habitable zones 
of main-sequence stars. On a longer time scale, infrared space interferometers 
could be built to directly image potentially habitable Earth-like planets and 
to search for life signatures in their atmospheres.

In summary, Doppler spectroscopy will continue to serve as a front-line 
detection technique for the next years. It will additionally provide an
increasingly important support to alternative planet search techniques, in
particular transit photometry and astrometry. The golden age of radial 
velocities is thus far from its end and Doppler spectroscopy promises 
additional exciting results to come.

\bibliographystyle{aa}
\bibliography{EggenbergerUdry}

\end{document}